\documentstyle[epsfig]{article}
\begin{document}

\centerline{ \Large \bf Introduction to the Random Matrix Theory:}
\centerline{\large \bf Gaussian Unitary Ensemble and Beyond}

\centerline{}

\vskip 0.4cm \centerline{\large \bf Yan V. Fyodorov}

\vskip 0.3cm

\centerline{Department of Mathematical Sciences, Brunel
University,} \centerline{Uxbridge, UB8 3PH, United Kingdom.}

\vskip 0.3cm

\begin{abstract}

These lectures provide an informal introduction into the notions
and tools used to analyze statistical properties of eigenvalues of
large random Hermitian matrices. After developing the general
machinery of orthogonal polynomial method,  we study in most
detail Gaussian Unitary Ensemble (GUE) as a paradigmatic example.
In particular, we discuss Plancherel-Rotach asymptotics of Hermite
polynomials in various regimes and employ it in spectral analysis
of the GUE. In the last part of the course we discuss general
relations between orthogonal polynomials and characteristic
polynomials of random matrices which is an active area of current
research.

\end{abstract}

\section{Preface}

Gaussian Ensembles of random Hermitian or real symmetric matrices
always played a prominent role in the development and applications
of Random Matrix Theory. Gaussian Ensembles are uniquely singled
out by the fact that they belong both to the family of invariant
ensembles, and to the family of ensembles with independent,
identically distributed (i.i.d) entries. In general, mathematical
methods used to treat those two families are very different.

In fact, all random matrix techniques and ideas can be most
clearly and consistently introduced using Gaussian case as a
paradigmatic example. In the present set of lectures we mainly
concentrate on consequences of the invariance of the corresponding
probability density function, leaving aside methods of exploiting
statistical independence of matrix entries. Under these
circumstances the method of orthogonal polynomials is the most
adequate one, and for the Gaussian case the relevant polynomials
are Hermite polynomials. Being mostly interested in the limit of
large matrix sizes we will spend a considerable amount of time
investigating various asymptotic regimes of Hermite polynomials,
since the latter are main building blocks of various correlation
functions of interest. In the last part of our lecture course we
will discuss why statistics of characteristic polynomials of
random Hermitian matrices turns out to be interesting and
informative to investigate, and will make a contact with recent
results in the domain.

The presentation is quite informal in the sense that I will not
try to prove various statements in full rigor or generality. I
rather attempt outlining the main concepts, ideas and techniques
preferring a good illuminating example to a general proof. A much
more rigorous and detailed exposition can be found in the cited
literature. I will also frequently employ the symbol $\propto$. In
the present set of lectures it always means that the expression
following $\propto$ contains a multiplicative constant which is of
secondary importance for our goals and can be restored when
necessary.

\section{Introduction}
In these lectures we use the symbol $^T$ to denote matrix or
vector transposition  and the asterisk $^*$ to denote Hermitian
conjugation. In the present section the bar $\overline{z}$ denotes
complex conjugation.

Let us start with a square complex matrix $\hat{Z}$ of dimensions
$N\times N$, with complex entries $z_{ij}=x_{ij}+iy_{ij},\, 1\le
i,j\le N$. Every such matrix can be conveniently looked at as a
point in a $2N^2$-dimensional Euclidean space with real Cartesian
coordinates $x_{ij},\,y_{ij}$, and the length element in this
space is defined in a standard way as:
\begin{equation}\label{1}
(ds)^2=\mbox{Tr}\left(d\hat{Z}d\hat{Z}^*\right)=\sum_{ij}dz_{ij}
\overline{dz_{ij}}=\sum_{ij}\left[(dx)^2_{ij}+(dy)^2_{ij}\right].
\end{equation}

As is well-known (see e.g.\cite{Fom}) any surface embedded in an
Euclidean space inherits a natural Riemannian metric from the
underlying Euclidean structure. Namely, let the coordinates in a
$n-$dimensional Euclidean space be $(x_1,\ldots,x_n)$, and let a
$k-$dimensional surface embedded in this space be parameterized in
terms of coordinates $(q_1,\ldots,q_k),\, k\le n$ as
$x_i=x_i(q_1,\ldots,q_k),\, i=1,\ldots n$. Then the Riemannian
metric $g_{ml}=g_{lm}$ {\it on the surface} is defined from the
Euclidean length element according to
\begin{equation}\label{2}
(ds)^2=\sum_{i=1}^n(dx_i)^2=\sum_{i=1}^n\left(\sum_{m=1}^k
\frac{\partial x_i}{\partial q_m}dq_m\right)^2 =\sum_{m,l=1}^k
g_{mn}dq_{m}dq_{l}.
\end{equation}
Moreover, such a Riemannian metric induces the corresponding integration
measure on the surface, with the volume element given by
\begin{equation}\label{3}
d\mu=\sqrt{|g|}dq_1\ldots dq_k,\quad
g=\det{\left(g_{ml}\right)_{l,m=1}^k}.
\end{equation}

For $k=n$ these are just the familiar formulae for the lengths and
volume associated with change of coordinates in an Euclidean
space. For example, for $n=2$ we can pass from Cartesian
coordinates $-\infty<x,\, y<\infty $ to polar coordinates $r>0$,
$0\le \theta<2\pi$ by $x=r\cos{\theta}$, $y=r\sin{\theta}$, so
that $dx=dr\cos{\theta}-r\sin{\theta}d\theta$, $dy=
dr\sin{\theta}+r\cos{\theta}d\theta$, and the Riemannian metric is
defined by $(ds)^2=(dx)^2+(dy)^2=(dr)^2+r^2(d\theta)^2$. We find
that $g_{11}=1, \, g_{12}=g_{21}=0, \, g_{22}=r^2$, and the volume
element of the integration measure in the new coordinates is
$d\mu=rdrd\theta$; as it should be. As the simplest example of a
``surface" with $k<n=2$ embedded in such a two-dimensional space
we consider a circle $r=R=const$. We immediately see that the
length element $(ds)^2$ restricted to this ``surface" is
$(ds)^2=R^2(d\theta)^2$, so that $g_{11}=R^2$, and the integration
measure induced on the surface is correspondingly $d\mu=Rd\theta$.
The ``surface" integration then gives the total ``volume" of the
embedded surface (i.e. circle length $2\pi R$).

\begin{figure}[h!]
\centering\epsfig{file=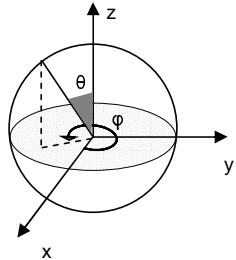, width=3.5cm} \caption{The
spherical coordinates for a two dimensional sphere in the
three-dimensional Euclidean space.} \label{fig1}
\end{figure}

Similarly, we can consider a two-dimensional ($k=2$) sphere
$R^2=x^2+y^2+z^2$ embedded in a three-dimensional Euclidean space
$(n=3)$ with coordinates $x,y,z$ and length element
$(ds)^2=(dx)^2+(dy)^2+(dz)^2$.
 A natural parameterization of the points on the sphere is possible
in terms of the spherical coordinates $\phi,\theta$ (see Fig.~
\ref{fig1})
\[
x=R\sin{\theta}\cos{\phi},\,\, y=R\sin{\theta}\sin{\phi},\,\,
z=R\cos{\theta};\quad 0\le \theta\le\pi,\,\, 0\le \phi < 2\pi,
\]
which results in
$(ds)^2=R^2(d\theta)^2+R^2\sin^2{\theta}(d\phi)^2$. Hence the
matrix elements of the metric are $g_{11}=R^2,\,g_{12}=g_{21}=0,
g_{22}=R^2\sin^2{\theta}$, and the corresponding ``volume element"
on the sphere is the familiar elementary area
$d\mu=R^2\sin{\theta}d\theta d\phi$.

As a less trivial example to be used later on consider a
$2-$dimensional manifold formed by $2\times 2$ unitary matrices
$\hat{U}$ embedded in the $8$ dimensional Euclidean space of
$Gl(2;C)$ matrices. Every such matrix can be represented as the
product of a matrix $\hat{U}_c$ from the coset space
$U(2)/U(1)\times U(1)$ parameterized by $k=2$ real coordinates
$0\le \phi<2\pi,\, 0\le \theta\le \pi/2$, and a diagonal unitary
matrix $U_d$, that is $\hat{U}=\hat{U}_d\hat{U}_c$, where
\begin{equation}\label{4}
\hat{U_c}=\left(\begin{array}{cc}
\cos{\theta}&-\sin{\theta}e^{-i\phi}\\
\sin{\theta}e^{i\phi}&\cos{\theta}\end{array}\right),\quad
\hat{U_d}=\left(\begin{array}{cc} e^{-i\phi_1}&0\\
0&e^{i\phi_2}\end{array}\right).
\end{equation}
Then the differential $d\hat{U}$
of the matrix $\hat{U}=\hat{U}_d\hat{U}_c$ has the following form:
\begin{equation}\label{5}
\hat{dU}=\left(\begin{array}{cc}
-[d\theta \sin{\theta}+i\cos{\theta}d\phi_1]e^{-i\phi_1}
& e^{-i(\phi_1+\phi)}[-d\theta\cos{\theta}+i(d\phi_1+d\phi)\sin{\theta}]
\\ e^{i(\phi+\phi_2)}
[d\theta\cos{\theta}+i(d\phi+d\phi_2)\sin{\theta}]
& [-d\theta \sin{\theta}+id\phi_2\cos{\theta}] e^{i\phi_2}
\end{array}\right),
\end{equation}
which yields the length element and the induced Riemannian metric:
\begin{equation}\label{6}
(ds)^2=\mbox{Tr}\left(d\hat{U}d\hat{U}^*\right)=
2(d\theta)^2+(d\phi_1)^2+(d\phi_2)^2+
2\sin^2{\theta}(d\phi)^2+2\sin^2{\theta}(d\phi\, d\phi_1 +d\phi\,
d\phi_2).
\end{equation}
We see that the nonzero entries of the Riemannian metric tensor
$g_{mn}$ in this case are
 $g_{11}=2,\, g_{22}=g_{33}=1, g_{44}=2\sin^2{\theta}, \,
 g_{24}=g_{42}=g_{34}=g_{43}=\sin^2{\theta}$, so that the
determinant $\det{[g_{mn}]}=4\sin^2{\theta}\cos^2{\theta}$.
Finally, the induced integration measure on the group $U(2)$ is
given by
\begin{equation}\label{7}
d\mu(\hat{U})=2\sin{\theta}\cos{\theta}\,d\theta\, d\phi\,
d\phi_1\, d\phi_2.
\end{equation}
It is immediately clear that the above expression is invariant, by
construction, with respect to multiplications $\hat{U}\to
\hat{V}\hat{U}$, for any fixed unitary matrix $V$ from the same
group. Therefore, Eq.(\ref{7}) is just the Haar measure on the
group.

We will make use of these ideas several times in our lectures. Let
us now concentrate on the $N^2-$dimensional subspace of Hermitian
matrices in the $2N^2-$ dimensional space of all complex matrices
of a given size $N$. The Hermiticity condition
$\hat{H}=\hat{H}^*\equiv\overline{\hat{H}^T}$ amounts to imposing
the following restrictions on the coordinates:
$x_{ij}=x_{ji},\,\,y_{ij}=-y_{ji}$. Such a restriction from the
space of general complex matrices results in the length and volume
element on the subspace of Hermitian matrices:
\begin{equation}\label{8}
(ds)^2=\mbox{Tr}\left(d\hat{H}d\hat{H}^*\right)=
\sum_{i}(dx_{ii})^2+2\sum_{i<j}\left[(dx_{ij})^2+(dy_{ij})^2\right]
\end{equation}
\begin{equation}\label{8a}
 d\mu(\hat{H})=2^{\frac{N(N-1)}{2}}\prod_{i}dx_{ii}\prod_{i<j}dx_{ij}dy_{ij}.
\end{equation}
Obviously, the length element $(ds)^2=\mbox{Tr}d\hat{H}d\hat{H}^*$
is invariant with respect to an automorphism (a mapping of the
space of Hermitian matrices to itself) by a similarity
transformation $\hat{H}\to U^{-1}\hat{H}\hat{U}$, where
$\hat{U}\in U(N)$ is any given unitary $N\times N$ matrix:
$\hat{U}^*=\hat{U}^{-1}$. Therefore the corresponding integration
measure $d\mu(\hat{H})$ is also invariant with respect to all such
``rotations of the basis".

The above-given measure $d\mu(\hat{H})$ written in the coordinates
$x_{ii},\, x_{i<j},\, y_{i<j}$ is frequently referred to as the
``flat measure". Let us discuss now another, very important
coordinate system in the space of Hermitian matrices which will be
in the heart of all subsequent discussions. As is well-known,
every Hermitian matrix $\hat{H}$ can be represented as
\begin{equation}\label{81}
\hat{H}=\hat{U}\hat{\Lambda}\hat{U}^{-1},\quad
\hat{\Lambda}=\mbox{diag} (\lambda_1,\ldots,\lambda_N),\,\,
\hat{U}^*\hat{U}=\hat{I},
\end{equation}
where real $-\infty<\lambda_k<\infty,\,k=1,\ldots,N$ are
eigenvalues of the Hermitian matrix, and rows of the unitary
matrix $\hat{U}$ are corresponding eigenvectors. Generically, we
can consider all eigenvalues to be simple (non-degenerate). More
precisely, the set of matrices $\hat{H}$ with non-degenerate
eigenvalues is dense and open in the $N^2$-dimensional space of
all Hermitian matrices, and has full measure (see \cite{Deift},
p.94 for a formal proof). The correspondence $\hat{H}\to
\left(\hat{U}\in U(N),\hat{\Lambda}\right)$ is, however, not
one-to-one, namely
$\hat{U}_1\hat{\Lambda}\hat{U}_1^{-1}=\hat{U}_2\hat{\Lambda}\hat{U}_2^{-1}$
if
$\hat{U}^{-1}_1\hat{U}_2=\mbox{diag}\left(e^{i\phi_1},\ldots,e^{i\phi_N}\right)$
for any choice of the phases $\phi_1,\ldots,\phi_N$. To make the
correspondence one-to-one we therefore have to restrict the
unitary matrices to the coset space $U(N)/U(1)
\otimes\ldots\otimes U(1)$, and also to order the eigenvalues,
e.g. requiring $\lambda_1<\lambda_2<\ldots<\lambda_N$. Our next
task is to write the integration measure $d\mu(\hat{H})$ in terms
of eigenvalues $\hat{\Lambda}$ and matrices $\hat{U}$. To this
end, we differentiate the spectral decomposition
$\hat{H}=\hat{U}\hat{\Lambda}\hat{U}^{*}$, and further exploit:
$d\left(\hat{U}^*\hat{U}\right)=d\hat{U}^*\hat{U}+\hat{U}^*d\hat{U}=0$.
This leads to
\begin{equation}\label{82}
d\hat{H}=\hat{U}\left[d\hat{\Lambda}+\hat{U}^{*}d\hat{U}\hat{\Lambda}
-\hat{\Lambda}\hat{U}^{*}d\hat{U}\right]\hat{U}^{*}.
\end{equation}
Substituting this expression into the length element $(ds)^2$, see
Eq.(\ref{8}), and using the short-hand notation $\delta\hat{U}$
for the matrix $\hat{U}^*d\hat{U}$ satisfying anti-Hermiticity
condition $\delta\hat{U}^*=-\delta\hat{U}$, we arrive at:
\begin{equation}\label{83}
(ds)^2=\mbox{Tr}\left[\left(d\hat{\Lambda}\right)^2+2d\hat{\Lambda}
\left(\delta\hat{U}\hat{\Lambda}
-\hat{\Lambda}\delta\hat{U}\right)+\left(\delta\hat{U}\hat{\Lambda}\right)^2
+\left(\hat{\Lambda}\delta\hat{U}\right)^2-2\delta\hat{U}
\hat{\Lambda}^2\delta\hat{U}\right].
\end{equation}
Taking into account that $\hat{\Lambda}$ is purely diagonal, and
therefore the diagonal entries of the commutator
$\left(\delta\hat{U}\hat{\Lambda}
-\hat{\Lambda}\delta\hat{U}\right)$ are zero, we see that the
second term in Eq.(\ref{83}) vanishes. On the other hand, the
third and subsequent terms when added up are equal to
\[
2\mbox{Tr}\left[\delta\hat{U}\hat{\Lambda}\delta\hat{U}\hat{\Lambda}-
\delta\hat{U}^2\hat{\Lambda}^2\right]=2\sum_{ij}\left[
\delta U_{ij}\lambda_j\delta U_{ji}\lambda_i-\lambda_i^2
\delta U_{ij}\delta U_{ji}\right]=-\sum_{ij}\left(\lambda_i-
\lambda_j\right)^2\delta U_{ji}\delta U_{ij}\,,
\]
which together with the first term yields the final expression for
the length element in the ``spectral" coordinates
\begin{equation}\label{84}
(ds)^2=\sum_{i}\left(d\lambda_i\right)^2+\sum_{i<j}\left(\lambda_i-
\lambda_j\right)^2\overline{\delta U_{ij}}\delta U_{ij}\,
\end{equation}
where we exploited the anti-Hermiticity condition $-\delta U_{ji}=
\overline{\delta U_{ij}}$. Introducing the real and imaginary
parts $\delta U_{ij}=\delta p_{ij}+i\delta q_{ij}$ as independent
coordinates we can calculate the corresponding integration measure
$d\mu(\hat{H})$ according to the rule in Eq.(\ref{3}), to see that
it is given by
\begin{equation}\label{85}
d\mu(\hat{H})=\prod_{i<j}\left(\lambda_i-
\lambda_j\right)^2\prod_{i}d\lambda_i \times d{\cal M}(\hat{U})\,.
\end{equation}
The last factor $d{\cal M}(U)$ stands for the part of the measure
depending only on the $U-$variables. A more detailed consideration
shows that, in fact, $d{\cal M}(\hat{U})\equiv d\mu(\hat{U})$,
which means that it is given (up to a constant factor) by the
invariant Haar measure on the unitary group $U(N)$. This fact is
however of secondary importance for the goals of the present
lecture.

Having an integration measure at our disposal, we can introduce a
probability density function (p.d.f.) ${\cal P}(\hat{H})$ on the
space of Hermitian matrices, so that ${\cal P}(\hat{H})
d\mu(\hat{H})$ is the probability that a matrix $\hat{H}$ belongs
to the volume element $d\mu(\hat{H})$. Then it seems natural to
require for such a probability to be invariant with respect to all
the above automorphisms, i.e. ${\cal P}(\hat{H})={\cal
P}\left(\hat{U}^*\hat{H}\hat{U}\right)$. It is easy to understand
that this ``postulate of invariance" results in ${\cal P}$ being a
function of $N$ first traces $\mbox{Tr}\hat{H}^n,\, n=1,\ldots,N$
(the knowledge of first $N$ traces fixes the coefficients of the
characteristic polynomial of $\hat{H}$ uniquely, and hence the
eigenvalues. Therefore traces of higher order can always be
expressed in terms of the lower ones).  Of particular interest is
the case
\begin{equation}\label{9}
{\cal P}(\hat{H})=C\exp{-Tr\,Q(\hat{H})},\quad
Q(x)=a_{2j}x^{2j}+\ldots +a_0,
\end{equation}
where $2j\le N$, the parameters $a_{2l}$ and $C$ are real constants,
and $a_{2j}>0$.
Observe that if we take
\begin{equation}\label{10}
Q(x)=ax^{2}+bx+c,
\end{equation}
then $e^{-Tr\,Q(\hat{H})}$ takes the form of the product
\begin{equation}\label{11}
e^{-a\left[\sum_ix^2_{ii}+2
\sum_{i<j}(x^2_{ij}+y^2_{ij})\right]}e^{-b\sum_ix_{ii}}e^{-cN}
=e^{-cN}\prod_{i=1}^N\left( e^{-ax_{ii}^2-bx_{ii}}\right)
\prod_{i<j}e^{-2ax_{ij}^2}\prod_{i<j}e^{-2ay_{ij}^2}.
\end{equation}
We therefore see that the probability distribution of the matrix
$\hat{H}$ can be represented as a product of factors, each factor
being a suitable Gaussian distribution depending only on one
variable in the set of all coordinates $x_{ii},\, x_{i<j},\,
y_{i<j}$. Since the same factorization is valid also for the
integration measure $d\mu(\hat{H})$, see Eq.(\ref{8a}), we
conclude that all these $N^2$ variables are statistically
independent and Gaussian-distributed.

A much less obvious statement is that if we impose {\it simultaneously}
two requirements:
\begin{itemize}
\item The probability density function ${\cal P}(\hat{H})$ is
invariant with respect to all conjugations $\hat{H}\to
\hat{H}'=U^{-1}\hat{H}\hat{U}$ by unitary matrices $\hat{U}$, that
is ${\cal P}(\hat{H}')={\cal P}(\hat{H})$; and \item the $N^2$
variables $x_{ii},\, x_{i<j},\, y_{i<j}$ are statistically
independent, i.e.
\begin{equation}\label{11a}
{\cal P}(\hat{H})=\prod_{i=1}^Nf_i(x_{ii})
\prod_{i<j}^Nf^{(1)}_{ij}(x_{ij})f^{(2)}_{ij}(y_{ij}),
\end{equation}
\end{itemize}
then the function ${\cal P}(\hat{H})$ is {\it necessarily} of the
form ${\cal P}(\hat{H})=Ce^{-\left(a\mbox{Tr}\hat{H}^2+b\mbox{Tr}
\hat{H}+cN\right)}$, for some constants $a>0,\, b,\, c$. The proof
for any $N$ can be found in \cite{Mehta}, and here we just
illustrate its main ideas for the simplest, yet nontrivial case
$N=2$. We require invariance of the distribution with respect to
the conjugation of $\hat{H}$ by $\hat{U}\in U(2)$, and first
consider a particular choice of the unitary matrix $\hat{U}=
\left(\begin{array}{cc} 1&-\theta\\ \theta&1\end{array}\right)$
corresponding to $\phi=\phi_1=\phi_2=0$, and small values $\theta
\ll 1$ in Eq.(\ref{4}). In this approximation the condition
$\hat{H}'=U^{-1}\hat{H}\hat{U}$ amounts to
\begin{equation}\label{11b}
\left(\begin{array}{cc}
x'_{11}& x'_{12}+iy'_{12} \\ x'_{12}-iy'_{12}&
x'_{22}\end{array}\right)=\left(\begin{array}{cc}
x_{11}+2\theta x_{12}&
x_{12}+iy_{12}+\theta\left(x_{22}-x_{11}\right)
\\ x_{12}-iy_{12}+\theta\left(x_{22}-x_{11}\right)&
x_{22}-2\theta x_{12}\end{array}\right),
\end{equation}
where we kept only terms linear in $\theta$. With the same precision
we expand the factors in Eq.(\ref{11a}):
\[
f_1(x'_1)=f_1(x_1)\left[1+2\theta x_{12}\frac{1}{f_1}
\frac{df_1}{dx_{11}}\right],\,\,\,
f_2(x'_{22})=f_2(x_{22})\left[1-2\theta x_{12}\frac{1}{f_2}
\frac{df_2}{dx_{22}}\right]
\]
\[
f^{(1)}_{12}(x'_{12})=f^{(1)}_{12}(x_{12})
\left[1+\theta (x_{22}-x_{11})\frac{1}{f^{(1)}_{12}}
\frac{df^{(1)}_{12}}{dx_{12}}\right],\,\,\,
f^{(2)}_{12}(y'_{21})=f^{(2)}_{12}(y_{12}).
\]
The requirements of statistical independence and invariance
amount to the product of the left-hand sides of the above expressions
to be equal to the product of the right-hand sides, for any $\theta$.
This is possible only if:
\begin{equation}\label{11c}
2x_{12}\left[\frac{d\ln{f_1}}{dx_{11}}-\frac{d\ln{f_2}}{dx_{22}}\right]
+(x_{22}-x_{11})\frac{d\ln{f^{(1)}_{12}}}{dx_{12}}=0,
\end{equation}
which can be further rewritten as
\begin{equation}\label{11d}
\frac{1}{(x_{22}-x_{11})}
\left[\frac{d\ln{f_1}}{dx_{11}}-\frac{d\ln{f_2}}{dx_{22}}\right]=const=
\frac{1}{2x_{12}}\frac{d\ln{f^{(1)}_{12}}}{dx_{12}},
\end{equation}
where we used that the two sides in the equation above depend on
essentially different sets of variables. Denoting $const_1=-2a$,
we see immediately that
\[
f^{(1)}_{12}(x_{12})\propto e^{-2ax^2_{12}},
\]
and further notice that
\[
\frac{d\ln{f_1}}{dx_{11}}+2ax_{11}=const_2=\frac{d\ln{f_2}}{dx_{22}}+2ax_{22}
\]
by the same reasoning. Denoting $const_2=-b$, we find:
\begin{equation}\label{11e}
f_{1}(x_{11})\propto e^{-a x^2_{11}-bx_{11}},\quad
f_{2}(x_{22})\propto e^{-a x^2_{22}-bx_{22}},
\end{equation}
and thus we are able to reproduce the first two factors in
Eq.(\ref{11}). To reproduce the remaining factors we consider the
conjugation by the unitary matrix $
\hat{U_d}=\left(\begin{array}{cc} 1-i\alpha&0\\
0&1+i\alpha\end{array}\right)$, which corresponds to the choice
$\theta=0,\,\phi_1=\phi_2=-\alpha=$ in Eq.(\ref{4}), and again we
keep only terms linear in the small parameter $\alpha\ll 1$.
Within such a precision the transformation leaves the diagonal
entries $x_{11},\, x_{22}$ unchanged, whereas the real and
imaginary parts of the off-diagonal entries are transformed as
\[
x'_{12}=x_{12}-2\alpha y_{12},\quad y'_{12}=y_{12}+2\alpha x_{12}.
\]
In this case the invariance of the p.d.f. ${\cal P}(\hat{H})$
together with the statistical independence of the entries amount,
after straightforward manipulations, to the condition
\[
\frac{1}{x_{12}}\frac{d\ln{f^{(1)}_{12}}}{dx_{12}}
=\frac{1}{y_{12}}\frac{d\ln{f^{(2)}_{12}}}{dy_{12}}
\]
which together with the previously found $f^{(1)}_{12}(x_{12})$
yields
\[
f^{(2)}_{12}(y_{12})\propto e^{-2ay^2_{12}},
\]
completing the proof of Eq.(\ref{11}).

The Gaussian form of the probability density function,
Eq.(\ref{11}), can also be found as a result of rather different
lines of thought.  For example, one may invoke an information
theory approach {\it a la} Shanon-Khinchin and define the amount
of information ${\cal I} [{\cal P}(\hat{H})]$ associated with any
probability density function ${\cal P}(\hat{H})$ by
\begin{equation}\label{12}
{\cal I} [{\cal P}(\hat{H})]=-\int d\mu(\hat{H})\,{\cal P}(\hat{H})
\ln{{\cal P}(\hat{H})}
\end{equation}
This is a natural extension of the corresponding
definition ${\cal I} [p_1,\ldots,p_m]=-\sum_{l=1}^mp_m\ln{p_m}$
for discrete events $1,...,m$.

Now one can argue that in order to have matrices $\hat{H}$ as
random as possible one has to find the p.d.f. minimizing the
information associated with it for a certain class of ${\cal
P}(H)$ satisfying some conditions. The conditions usually have a
form of constraints ensuring that the probability density function
has desirable properties.  Let us, for example, impose the only
requirement that the ensemble average for the two lowest traces
$\mbox{Tr}\hat{H},\mbox{Tr}\hat{H}^2$ must be equal to certain
prescribed values, say $E\left[\mbox{Tr}\hat{H}\right]=b$ and
$E\left[\mbox{Tr}\hat{H}^2\right]=a>0$, where the $E\left[\ldots
\right]$ stand for the expectation value with respect to the
p.d.f. ${\cal P}(H)$. Incorporating these constraints into the
minimization procedure in a form of Lagrange multipliers
$\nu_1,\nu_2$, we seek to minimize the functional
\begin{equation}\label{13}
{\cal I} [{\cal P}(\hat{H})]=-\int d\mu(\hat{H})\,{\cal
P}(\hat{H}) \left\{\ln{{\cal P}(\hat{H})}-\nu_1
\mbox{Tr}\hat{H}-\nu_2\mbox{Tr}\hat{H}^2\right\}.
\end{equation}
The variation of such a functional with respect to $\delta{\cal
P}(\cal{H})$ results in
\begin{equation}\label{14}
\delta{\cal I} [{\cal P}(\hat{H})]=-\int
d\mu(\hat{H})\,\delta{\cal P}(\hat{H}) \left\{1+\ln{{\cal
P}(\hat{H})}-\nu_1
\mbox{Tr}\hat{H}-\nu_2\mbox{Tr}\hat{H}^2\right\}=0
\end{equation}
possible only if \[{\cal P}(\hat{H})\propto\exp\{\nu_1
\mbox{Tr}\hat{H}+\nu_2\mbox{Tr}\hat{H}^2\}\] again giving the
Gaussian form of the p.d.f. The values of the Lagrange multipliers
are then uniquely fixed by constants $a,b$, and the normalization
condition on the probability density function. For more detailed
discussion, and for further reference see \cite{Mehta}, p.68.

Finally, let us discuss yet another construction allowing one to
arrive at the Gaussian Ensembles exploiting the idea of Brownian
motion. To start with, consider a system whose state at time $t$
is described by one real variable $x$, evolving in time according
to the simplest linear differential equation $\frac{d}{dt}x=-x$
describing a simple exponential relaxation $x(t)=x_0e^{-t}$
towards the stable equilibrium $x=0$. Suppose now that the system
is subject to a random additive Gaussian white noise $\xi(t)$
function of intensity $D$ \footnote{The following informal but
instructive definition of the white noise process may be helpful
for those not very familiar with theory of stochastic processes.
For any positive $t>0$ and integer $k\ge 1$ define the random
function $\xi_k(t)=\sqrt{2/\pi}\sum_{n=0}^k a_n\cos{nt}$, where
real coefficients $a_n$ are all independent, Gaussian distributed
with zero mean $E[a_n]=0$ and variances $E[a^2_0]=D/2$ and
$E[a^2_n]=D$ for $1\le n\le k$. Then one can, in a certain sense,
consider white noise as the limit of $\xi_k(t)$ for $k\to \infty$.
In particular, the Dirac $\delta(t-t')$ is approximated by the
limiting value of $\frac{\sin{[(k+1/2)(t-t')]}}{2\pi
\sin{(t-t')/2}}$}, so that the corresponding equation acquires the
form
\begin{equation}\label{Br1}
\frac{d}{dt}x=-x+\xi(t),\quad E_{\xi}\left[
\xi(t_1)\xi(t_1)\right]=D\delta(t_1-t_2),
\end{equation}
where $E_{\xi}[\ldots]$ stands for the expectation value with
respect to the random noise. The main characteristic property of a
Gaussian white noise process is the following identity:
\begin{equation}\label{Br2}
E_{\xi}\left[\exp\left\{\int_a^b\xi(t)v(t)dt\right\}\right]=
\exp\left\{\frac{D}{2}\int_a^bv^2(t)dt\right\}
\end{equation}
valid for any (smooth enough) test function $v(t)$. This is just a
direct generalization of the standard Gaussian integral identity:
\begin{equation}\label{Gauint0}
\int_{-\infty}^{\infty}
\,\frac{dq}{\sqrt{2\pi a}}\, e^{-\frac{1}{2a}q^2+qb}=e^{\frac{ab^2}{2}}.
\end{equation}
valid for $\mbox{Re}\,a>0$, and any (also complex) parameter $b$.

For any given realization of the Gaussian random process $\xi(t)$
the solution of the stochastic differential equation
Eq.(\ref{Br1}) is obviously given by
\begin{equation}\label{Br3}
x(t)=e^{-t}\left[x_0+\int_{0}^te^{\tau}\xi(\tau)d\tau\right].
\end{equation}
This is a random function, and our main goal is to find the
probability density function ${\cal P}(t,x)$ for the variable
$x(t)$ to take value $x$ at any given moment in time $t$, if we
know surely that $x(0)=x_0$. This p.d.f. can be easily found from
the characteristic function
\begin{equation}\label{Br4}
{\cal F}(t,q)= E_{\xi}\left[e^{-iqx(t)}\right]
=\exp\left\{-iqx_0e^{-t}-\frac{Dq^2}{4}(1-e^{-2t})\right\}
\end{equation}
obtained by using Eqs. (\ref{Br2}) and (\ref{Br3}).
The p.d.f. is immediately recovered by employing the inverse
Fourier transform:
\begin{equation}\label{Br5}
{\cal P}(t,x)=\int_{-\infty}^{\infty} \frac{dq}{2\pi} e^{iqx}
E_{\xi}\left[e^{-iqx(t)}\right] =\frac{1}{\sqrt{\pi D(1-e^{-2t})}}
\exp\left\{-\frac{\left(x-x_0e^{-t}\right)^2}{D(1-e^{-2t})}\right\}.
\end{equation}

The formula Eq.(\ref{Br5}) is called the Ornstein-Uhlenbeck (OU)
probability density function, and the function $x(t)$ satisfying
the equation Eq.(\ref{Br1}) is known as the O-U process.  In fact,
such a process describes an interplay between the random ``kicks"
forcing the system to perform a kind of Brownian motion and the
relaxation towards $x=0$. It is easy to see that when time grows
the OU p.d.f. ``forgets" about the initial state and tends to a
{\it stationary} (i.e. time-independent) universal Gaussian
distribution:
\begin{equation}\label{Br6}
{\cal P}(t\to\infty,x)=\frac{1}{\sqrt{\pi D}}
\exp\left\{-\frac{x^2}{D}\right\}.
\end{equation}

Coming back to our main topic, let us consider $N^2$ independent OU processes:
$N$ of them denoted as
\begin{equation}\label{Br7}
\frac{d}{dt}x_{i}=-x_{i}+\xi_i(t),\quad 1\le i\le N
\end{equation}
and the rest $N(N-1)$ given by
\begin{equation}\label{Br8}
\frac{d}{dt}x_{ij}=-x_{ij}+\xi^{(1)}_{ij}(t), \quad
\frac{d}{dt}y_{ij}=-y_{ij}+\xi^{(2)}_{ij}(t),
\end{equation}
where the indices satisfy $\quad 1\le i<j\le N$. Stochastic
processes $\xi(t)$ in the above equations are taken to be all
mutually independent Gaussian white noises characterized by the
correlation functions:
\begin{equation}\label{Br9}
E_{\xi}\left[\xi_{i_1}(t_1)\xi_{i_2}(t_2)\right]=2D\delta_{i_1,i_2}
\delta(t_1-t_2),\,\,E_{\xi}\left[\xi^{\sigma_1}_{ij}(t_1)
\xi^{\sigma_2}_{kl}(t_2)\right]=
D\delta_{\sigma_1,\sigma_2}\delta_{i,k}\delta_{j,l}\delta(t_1-t_2).
\end{equation}
As initial values $x_i(0),x_{ij}(0),y_{ij}(0)$ for each OU process
we choose diagonal and off-diagonal entries $H_{ii}^{(0)},\,\,
i=1,\ldots,N$ and $\mbox{Re}H_{i<j}^{(0)},\,
\mbox{Im}H_{i<j}^{(0)}$ of a fixed $N\times N$ Hermitian matrix
$\hat{H}^{(0)}$. Let us now consider the Hermitian matrix
$\hat{H}(t)$ whose entries are $H_{ii}(t)=x_i(t),\,
H_{i<j}(t)=x_{i<j}(t)+iy_{i<j}(t)$ for any $t\ge 0$. It is
immediately clear that the joint p.d.f. ${\cal
P}\left(t,\hat{H}\right)$ of the entries of such a matrix
$\hat{H}(t)$  will be given for any $t\ge 0 $ by the OU-type
formula:
\begin{equation}\label{Br10}
{\cal P}(t,\hat{H}) \propto Const\times
\frac{1}{\sqrt{(1-e^{-2t})^{N^2}}}
\exp\left\{-\frac{1}{D(1-e^{-2t})}\mbox{Tr}\left(\hat{H}-\hat{H}_0
e^{-t}\right)^2\right\}.
\end{equation}
In the limit $t\to\infty$ this p.d.f. converges to a stationary,
$t-$independent expression
\begin{equation}\label{Br11}
{\cal P}(t,\hat{H})
\propto C\, e^{-\frac{1}{D}\mbox{Tr}\hat{H}^2}
\end{equation}
{\it independent} of the initial matrix $\hat{H}_0$. We see
therefore that the familiar Gaussian ensemble in the space of
Hermitian matrices arises as the result of the stochastic
relaxation from any initial condition, in particular, from any
diagonal matrix with uncorrelated entries. In the next step one
may try to deduce the stochastic dynamics of the {\it eigenvalues}
of the corresponding matrices. Those eigenvalues obviously evolve
from completely uncorrelated to highly correlated patterns.  This
very interesting set of question goes beyond our present goals and
we refer to \cite{Mehta} for an introduction into the problem.

After specifying the probability density function ${\cal P}(H)$
the main question of interest is to characterize the  statistical
properties of the sequence of eigenvalues
$\lambda_1,\ldots,\lambda_N$ of $\hat{H}$. A convenient way of
doing this is to start with the joint p.d.f. of all these
eigenvalues. Because of the ``rotational invariance" assumption
the function ${\cal P}(\hat{H})$ depends in fact only on the
eigenvalues, for example for the ``symmetric" Gaussian case $b=0$
we have ${\cal P}(\hat{H})\propto e^{-a\sum_{i=1}^N\lambda_i^2}$.
Moreover, we have seen  that the integration measure
$d\mu(\hat{H})$ when expressed in terms of eigenvalues and
eigenvectors effectively factorizes, see Eq.(\ref{85}). Collecting
all these facts we arrive at the conclusion, that the relevant
joint p.d.f of all eigenvalues can be always written, up to a
normalization constant, as
\begin{equation}\label{JPDG}
{\cal P}(\lambda_1,\ldots,\lambda_N)\,d\lambda_1 \ldots d\lambda_N
\propto e^{-\sum_{i=1}^NQ(\lambda_i)}
\prod_{i<j}\left(\lambda_i-\lambda_j\right)^2
\,\,d\lambda_1 \ldots d\lambda_N
\end{equation}
for a general, non-gaussian weight $e^{-\mbox{Tr}Q(\hat{H})}$. We
immediately see that the presence of the ``Jacobian factor"
$\prod_{i<j}\left(\lambda_i-\lambda_j\right)^2$ is responsible of
the fact that the eigenvalues are correlated in a non-trivial way.
In what follows we are going to disregard  the fact that
eigenvalues $\lambda_i$ were initially put in increasing order.
More precisely, for any {\it symmetric} function $f$  of $N$ real
variables $\lambda_1,\ldots,\lambda_N$ the expected value will be
calculated as
\[
\int_{{\bf R}^N} f(\lambda_1,\ldots,\lambda_N) {\cal
P}(\lambda_1,\ldots,\lambda_N)d\lambda_1\,\ldots\, d\lambda_N.
\]
Indeed, with p.d.f. being symmetric with respect to permutations
of the eigenvalue set, disregarding the ordering amounts to a
simple multiplicative combinatorial factor $n!$ in the
normalization constant.

Our main goal is to extract the information about these eigenvalue
correlations in the limit of large size $N$. From this point of
view it is pertinent to discuss a few quantitative measures
frequently used to characterize correlations in sequences of real
numbers.

\section{Characterization of Spectral Sequences}

Let $-\infty<\lambda_1,\lambda_2,\ldots,\lambda_N<\infty$ be the
positions of $N$ points on the real axis, characterized by the
joint probability density function (JPDF)
\[
{\cal P}(\lambda_1,\lambda_2,\ldots,\lambda_N)
\,\,d\lambda_1\,\ldots\,d\lambda_N
\]
of having, {\it regardless of labelling}, one point in the
interval $[\lambda_1,\lambda_1+d\lambda_1]$, another in the
interval $[\lambda_2,\lambda_2+d\lambda_2]$,..., another in
$[\lambda_N,\lambda_N+d\lambda_N]$. Since in this section we deal
exclusively with real variables, the bar will stand for the
expectation value with respect to such a JPDF.

The statistical properties of the sequence $\{\lambda_i\}$ are
conveniently characterized by the set of $n-$point {\it correlation
functions}, defined as
\begin{equation}\label{21}
{\cal R}_{n}(\lambda_1,\lambda_2,\ldots,\lambda_n)
=\frac{N!}{(N-n)!}\int{\cal
P}(\lambda_1,\lambda_2,\ldots,\lambda_N) \,\,
d\lambda_{n+1}\,\ldots\,d\lambda_N.
\end{equation}
It is obvious from this definition that the lower correlation
functions can be obtained from the higher-order ones:
\begin{equation}\label{22}
{\cal R}_{n}(\lambda_1,\lambda_2,\ldots,\lambda_n)
=\frac{1}{(N-n)}\int{\cal R}_{n+1}(\lambda_1,\lambda_2,\ldots,\lambda_{n+1})
\,\,  d\lambda_{n+1}.
\end{equation}

To provide a more clear interpretation of these correlation
functions we relate them to the statistics of the number $N_B$ of
points of the sequence $\{\lambda_i\}$ within any set $B$ of the
real axis (e.g an interval $[a,b]$). Let $\chi_B(x)$ be the
characteristic function of the set $B$, equal to unity if $x\in B$
and zero otherwise. Introduce the exact density function
$\rho_N(\lambda)$ of the points $\{\lambda_i\}$ around the point
$\lambda$ on the real axis. It can be conveniently written using
the Dirac's $\delta-$function as
$\rho_N(\lambda)=\sum_{i=1}^N\delta(\lambda-\lambda_i)$. Then
$N_B=\int\chi_B(\lambda)\rho_N(\lambda)d\lambda$.

On the other hand, consider
\begin{eqnarray}\label{23}
\nonumber &&\int_B{\cal R}_{1}(\lambda_1)d\lambda_1=
\int\chi_B(\lambda_1){\cal R}_{1}(\lambda_1)d\lambda_1
=N\int \chi_B(\lambda_1){\cal P}(\lambda_1,\lambda_2,\ldots,\lambda_N)
\,\, d\lambda_{1}\,\ldots\,d\lambda_N\\
 &&
=\int\, \sum_{i=1}^N\chi_B(\lambda_i){\cal P}
(\lambda_1,\lambda_2,\ldots,\lambda_N) \,\,
d\lambda_{1}\,\ldots\,d\lambda_N
\end{eqnarray}
and therefore
\begin{eqnarray}\label{24}
\int_B{\cal R}_{1}(\lambda_1)d\lambda_1=\overline{ N_B}=
\mbox{expectation of the number of points in B}.
\end{eqnarray}

Similarly, consider
\begin{eqnarray}\label{25}
\nonumber &&\int\chi_B(\lambda_1)\chi_B(\lambda_2) {\cal
R}_{2}(\lambda_1,\lambda_2)d\lambda_1d\lambda_2 =N(N-1)\int
\chi_B(\lambda_1)\chi_B(\lambda_2){\cal P}
(\lambda_1,\lambda_2,\ldots,\lambda_N)
\,\, d\lambda_{1}\,\ldots\,d\lambda_N\\
&& = \int\, \sum_{i\ne j}^N\chi_B(\lambda_i)\chi_B(\lambda_j)
{\cal P} (\lambda_1,\lambda_2,\ldots,\lambda_N) \,\,
d\lambda_{1}\,\ldots\,d\lambda_N,
\end{eqnarray}
which can be interpreted as
\begin{eqnarray}\label{26}
\int_{B\times B}{\cal R}_{2}(\lambda_1,\lambda_2)d\lambda_1d\lambda_2=
\mbox{expectation of the number of pairs of points in B}
\end{eqnarray}
where if, say, $\lambda_1$ and $\lambda_2$ are in $B$, then the pair
$\{1,2\}$ and $\{2,1\}$ are both counted.

To relate the two-point correlation function to the variance of
$N_B$ we notice that in view of Eq.(\ref{24}) the one-point
correlation function ${\cal R}_1(\lambda)$ coincides with the mean
density $\overline{\rho}_N(\lambda)$ of the points $\{\lambda_i\}$
around the point $\lambda$ on the real axis.  Similarly, write the
mean square $\overline{N^2_B}$
\begin{equation}\label{2p}
\overline{N^2_B}=\int\chi_B(\lambda)\chi_B(\lambda')
\overline{\rho_N(\lambda)\rho_N(\lambda')}\,\,d\lambda d\lambda'
\end{equation}
and notice that
\begin{eqnarray}\label{27}
\nonumber &&\overline{\rho_N(\lambda)\rho_N(\lambda')}=\overline{\sum_{ij}
\delta(\lambda-\lambda_i)\delta(\lambda'-\lambda_j)}
=\delta(\lambda-\lambda')\overline{\sum_{i}
\delta(\lambda-\lambda_i)}+\overline{\sum_{i\ne j}
\delta(\lambda-\lambda_i)\delta(\lambda'-\lambda_j)}\\
&& =\delta(\lambda-\lambda'){\cal R}_1(\lambda)+
{\cal R}_2(\lambda,\lambda').
\end{eqnarray}
In this way we arrive at the relation:
\begin{equation}
\overline{N^2_B}=\overline{N_B}+\int_{B\times B}
{\cal R}_{2}(\lambda,\lambda')\,\,d\lambda d\lambda'.
\end{equation}
In fact, it is natural to introduce the so-called ``number
variance" statistics
$\Sigma_2(B)=\overline{N^2_B}-\left[\overline{N_B}\right]^2$
describing the variance of the number of points of the sequence
inside the set  $B$. Obviously,
\begin{equation}\label{sigma2}
\Sigma_2(B)=\overline{N_B}+\int_{B\times B}\left[ {\cal
R}_{2}(\lambda,\lambda')-{\cal R}_{1}(\lambda){\cal
R}_{1}(\lambda') \right]\,\,d\lambda d\lambda'\equiv
\overline{N_B}-\int_{B\times B}
Y_{2}(\lambda,\lambda')\,\,d\lambda d\lambda'
\end{equation}
where we introduced the so-called {\it cluster function}
$Y_{2}(\lambda,\lambda')= {\cal R}_{1}(\lambda){\cal
R}_{1}(\lambda')-{\cal R}_{2}(\lambda,\lambda')$ frequently used
in applications.

Finally, in principle the knowledge of all $n-$point correlation
functions provides one with the possibility of calculating an
important characteristic of the spectrum known as the ``hole
probability" $A(L)$. This quantity is defined as the probability
for a random matrix to have {\it no} eigenvalues in the interval
$(-L/2,L/2)$ \footnote{Sometimes one uses instead the interval
$[-L,L]$ to define $A(L)$, see e.g. \cite{Deift}.}. Define
$\chi_L(\lambda)$ to be the characteristic function of this
interval. Obviously,
\begin{eqnarray}\label{hole}
&& A(L)=\int\ldots\int\, {\cal
P}(\lambda_1,\ldots,\lambda_N)\prod_{k=1}^N
\left(1-\chi_L(\lambda_k)\right)\,\, d\lambda_1 \ldots d\lambda_N\\
&& =\sum_{j=0}^N(-1)^j\, \int\ldots\int\, {\cal
P}(\lambda_1,\ldots,\lambda_N)h_j\left\{
\chi_L(\lambda_1),\ldots,\chi_L(\lambda_N)\right\}\,\, d\lambda_1
\ldots d\lambda_N,
\end{eqnarray}
where $h_j\{x_1,\ldots,x_N\}$ is the $j-th$ symmetric function:
\[
h_0\{x_1,\ldots,x_N\}=1,\,\,h_1\{x_1,\ldots,x_N\}=\sum_{i=1}^Nx_i\, ,
\]
\[
h_2\{x_1,\ldots,x_N\}=\sum_{i<j}^Nx_ix_j, \quad \ldots ,\quad
h_N\{x_1,\ldots,x_N\}=x_1x_2\ldots x_N.
\]
Now, for $1\le j\le N$
\begin{eqnarray}\label{hole1}
\nonumber && \int\ldots\int\,\prod_{k=1}^j \chi_L(\lambda_k)\,\,
{\cal
P}(\lambda_1,\ldots,\lambda_N) d\lambda_1 \ldots d\lambda_N\\
&&=\frac{(N-j)!}{N!} \int\ldots\int \prod_{k=1}^j
\chi_L(\lambda_k) {\cal R}_j(\lambda_1,\ldots,\lambda_j)\,\,
d\lambda_1 \ldots d\lambda_j\\&& =
\frac{(N-j)!}{N!}\int_{|x_1|<L/2}\ldots \int_{|x_j|<L/2}\, {\cal
R}_j(\lambda_1,\ldots,\lambda_j)\,\, d\lambda_1 \ldots d\lambda_j.
\end{eqnarray}
As $h_j$ contains $\left(\begin{array}{c}N \\ j\end{array}\right)$
terms and as ${\cal R}_j(\lambda_1,\ldots,\lambda_j)$ is invariant
under permutations of the arguments, it follows that
\begin{eqnarray}\label{hole2}
\nonumber && \int\ldots\int {\cal
P}(\lambda_1,\ldots,\lambda_N)h_j\left\{
\chi_L(\lambda_1)\ldots\chi_L(\lambda_N)\right\}\,\, d\lambda_1
\ldots d\lambda_j\\ && =\frac{(N-j)!}{N!}\left(\begin{array}{c}N
\\ j\end{array}\right)
\int_{|x_1|<L/2}\ldots \int_{|x_j|<L/2}\,
{\cal R}_j(\lambda_1,\ldots,\lambda_j)\,\,
d\lambda_1 \ldots d\lambda_j=\\ && \frac{1}{j!}
\int_{|x_1|<L/2}\ldots \int_{|x_j|<L/2}\,
{\cal R}_j(\lambda_1,\ldots,\lambda_j)\,\,
d\lambda_1 \ldots d\lambda_j.
\end{eqnarray}
Thus, we arrive at the following relation between the hole probability and
the $n-$point correlation functions:
\begin{eqnarray}\label{hole3}
&& A(L)=\sum_{j=0}^N\frac{(-1)^j}{j!}
\int_{-L/2}^{L/2}\ldots \int_{-L/2}^{L/2}\,
{\cal R}_j(\lambda_1,\ldots,\lambda_j)\,\,d\lambda_1 \ldots d\lambda_j.
\end{eqnarray}

One of the main goals of this set of lectures is to develop a
method allowing to evaluate all the $n-$point correlation
functions of the eigenvalues for any JPDF corresponding to unitary
invariant ensembles of the form Eq.(\ref{9}). After that we will
concentrate on a particular case of Gaussian weight and will
investigate the limiting behaviour of the kernel function
$K_n(\lambda,\lambda')$ as $N\to \infty$. But even before doing
this it is useful to keep in mind for reference purposes the
results corresponding to completely uncorrelated (a.k.a.
Poissonian) spectra. Those are described by a sequence of real
points $\lambda_1,\ldots,\lambda_N$, characterized by the fully
factorized JPDF:
\begin{equation}\label{uncor}
{\cal P}(\lambda_1,\lambda_2,\ldots,\lambda_N)
\,\,=p(\lambda_1)\,\ldots\,p(\lambda_N).
\end{equation}
The normalization condition requires  $\int_{-\infty}^{\infty}
p(\lambda)\,d\lambda=1$, and we further assume $p(\lambda)$ to be
a smooth enough integrable function. Obviously, for this case
\begin{equation}\label{uncor1}
{\cal R}_{n}(\lambda_1,\lambda_2,\ldots,\lambda_n)
=\frac{N!}{(N-n)!}p(\lambda_1)\,\ldots\,p(\lambda_n).
\end{equation}
In particular, ${\cal R}_{n}(\lambda)=Np(\lambda)$ which is just the
mean density $\overline{\rho(\lambda)}$ of points
$\{\lambda_i\}$ around the point $\lambda$ on
the real axis, and ${\cal
R}_{n}(\lambda_1,\lambda_2)=N(N-1)p(\lambda_1) p(\lambda_2)$,
etc.. From this we easily find for the number of levels in the
domain $B$ and for its mean square:
\begin{equation}\label{uncor2}
\overline{N_B}=N\int_{B}p(\lambda)\,d\lambda,\quad
\overline{N^2_B}=\overline{N_B}(N-1)/N+\left[\overline{N_B}\right]^2
\end{equation}
and for the hole probability
\begin{eqnarray}\label{uncorhole}
&& A(L)=\sum_{j=0}^N\frac{(-1)^j}{j!}\frac{N!}{(N-j)!}
\left[\int_{-L/2}^{L/2}  p(\lambda)\,d\lambda\right]^j=\left[1-
\int_{-L/2}^{L/2}  p(\lambda)\,d\lambda\right]^N.
\end{eqnarray}
Finally, let us specify $B$ to be the interval $[-L/2,L/2]$ around
the origin, and being interested mainly in large $N\gg 1$ consider
the length $L$ comparable with the mean spacing between
neighbouring points in the sequence $\{\lambda_i\}$ close to the
origin, given by $\Delta\equiv
\left[\overline{\rho(0)}\right]^{-1}= 1/[Np(0)]$. In other words
$s=L/\Delta=LNp(0)$ stays finite when $N\to\infty$. On the other
hand, for large enough $N$ the function $p(\lambda)$ can be
considered practically constant through the interval of the length
$L=O(1/N)$, and therefore the mean number of points of the
sequence $\{\lambda_i\}$ inside the interval $[-L/2,L/2]$ will be
asymptotically given by $\overline{N(s)}\approx N\,L\,p(0)=s$.
Similarly, using Eq.(\ref{uncor2}) one can easily calculate the
``number variance"
$\Sigma_2(s)=\overline{N^2_{[-\frac{L}{2},\frac{L}{2}]}} -
\left[\overline{N_{[-\frac{L}{2},\frac{L}{2}]}}\right]^2=
(N-1)\int_{-L/2}^{L/2} p(\lambda)\,d\lambda\approx s$. In the same
approximation the hole probability, Eq.(\ref{uncorhole}), tends
asymptotically to $A(s)\approx e^{-s}$. Later on we shall compare
these results with the corresponding behaviour emerging from the
random matrix calculations.

\section{The method of orthogonal polynomials}
In the heart of the method developed mainly by Dyson, Mehta and
Gaudin lies an ``integrating-out" Lemma \cite{Mehta}. In
presenting this material I follow very closely \cite{Deift},
pp.103-105.
\begin{itemize}
\item {\it
Let $J_n=J_n({\bf x})=(J_{ij})_{1\le i,j\le n}$ be an $n\times n$
matrix whose entries depend on a real vector ${\bf
x}=(x_1,x_2,\ldots,x_n)$ and have the form $J_{ij}=f(x_i,x_j)$,
where $f$ is some (in general, complex-valued) function satisfying
for some measure $d\mu(x)$ the ``reproducing kernel" property:
\begin{equation}\label{31}
\int f(x,y)f(y,z)\,d\mu(y)=f(x,z).
\end{equation}
Then
\begin{equation}\label{32}
\int \mbox{det}\,J_n({\bf x})\,d\mu(x_n)=[q-(n-1)]\mbox{det}\,J_{n-1}
\end{equation}
where $q=\int f(x,x)\,d\mu(x)$, and the matrix
$J_{n-1}=(J_{ij})_{1\le i,j\le n-1}$ have the same functional form
as $J_n$ with ${\bf x}$ replaced by $(x_1,x_2,\ldots,x_{n-1}).$ }
\end{itemize}

Before giving the idea of the proof for an arbitrary $n$ it is
instructive to have a look on the simplest case $n=2$, when
\[
J_2=\left(\begin{array}{cc}f(x_1,x_1)&f(x_1,x_2)\\
f(x_2,x_1)&f(x_2,x_2)\end{array} \right),\quad \mbox{hence}\quad
\det{J_n}=f(x_1,x_1)f(x_2,x_2)-f(x_1,x_2)f(x_2,x_1).
\]
Integrating the latter expression over $x_2$, and using the
``reproducing kernel" property, we immediately see that the result
is indeed just $(q-1)f(x,x)=(q-1)\det{J_1}$ in full agreement with
the statement of the Lemma.

For general $n$ one should follow essentially the same strategy and
expand the determinant as a sum over $n!$ permutations
$P_{n}(\sigma)=(\sigma_1,\ldots\sigma_n)$ of the index set $1,\ldots,n$
as
\begin{equation}\label{33}
\int \mbox{det}\,J_n({\bf x})\,d\mu(x_n)=\sum_{P_{n}}
(-1)^{P_n}\int f(x_1,x_{\sigma_1})\ldots f(x_n,x_{\sigma_n})
\,d\mu(x_n),
\end{equation}
where $(-1)^{P_n}$ stands for the sign of permutations. Now, we
classify the terms in the sum according to the actual value of the
index $\sigma_n=k,\,k=1,2,\ldots,n$. Consider first the case
$\sigma_n=n$, when effectively only the last factor $f(x_n,x_n)$
in the product is integrated yielding $d$ upon the integration.
Summing up over the remaining $(n-1)!$ permutations
$P_{n-1}(\sigma)$ of the index set $(1,2,...,n-1)$ we see that:
\[
\sum_{P_{n-1}}(-1)^{P_{n}}\int \, f(x_1,x_{\sigma_1})\ldots
f(x_n,x_{n}) \,d\mu(x_n) =q\sum_{P_{n-1}}(-1)^{P_{n-1}}\,
f(x_1,x_{\sigma_1})\ldots f(x_{n-1},x_{\sigma_{n-1}}),
\]
which is evidently equal to $q\det{J_{n-1}}$. Now consider
$(n-1)!$ terms with $\sigma_n=k<n$, when we have $\sigma_j=n$ for
some $j<n$. For every such term we have by the ``reproducing
property"
\[
\int \,f(x_1,x_{\sigma_1}) \ldots f(x_j,x_{n})\ldots
f(x_n,x_{k}) \,d\mu(x_n)
=f(x_1,x_{\sigma_1})\ldots f(x_j,x_{k})
\ldots f(x_{n-1},x_{\sigma_{n-1}}).
\]
Therefore
\begin{equation}\label{35}
\int \mbox{det}\,J_n({\bf x})\,d\mu(x_n)=
q\det{J_{n-1}}+\sum_{k=1}^{n-1}
\sum_{P_{n}:\sigma_n=k)}
(-1)^{P_n}f(x_1,x_{\sigma_1})\ldots f(x_j,x_k)
\ldots f(x_{n-1},x_{\sigma_{n-1}}).
\end{equation}
It is evident that the structure and the number of terms is as
required, and the remaining task is to show that the summation
over all possible $(n-1)!$ permutation of the index set for fixed
$k$ yields always $-\det{J_{n-1}}$, see \cite{Deift}. Then the
whole expression is indeed equal to $[q-(n-1)]\det{J_{n-1}}$ as
required.

Our next step is to apply this Lemma to calculating the $n-$point
correlation functions of the eigenvalues
$\lambda_1,\ldots,\lambda_n$ starting from the JPDF,
Eq.(\ref{JPDG}).

For this we notice that
\begin{equation}\label{vdm}
\prod_{i<j}^N(\lambda_i-\lambda_j)=(-1)^{\frac{N(N-1)}{2}}
\det{\left(\begin{array}{ccc}1&\ldots & 1\\ \lambda_1&\ldots & \lambda_N\\
.&.&.\\ .&.&.\\.&.&.\\\lambda^{N-1}_1&\ldots & \lambda^{N-1}_N
\end{array}\right)}\equiv \Delta_N(\lambda_1,\ldots,\lambda_N),
\end{equation}
where the determinant in the right-hand side is the famous van der
Monde determinant. Since the determinant cannot change upon
linearly combining its rows, the entries $\lambda_i^k$ in
$(k+1)-th$ row of the van der Monde determinant can be replaced,
up to a constant factor $a_0a_1...a_{N-1}$, by a polynomial of
degree $k$ of the form: $\pi_k(\lambda_i)=a_k\lambda_i^k+\mbox{any
polynomial in }\, \lambda_i\,
 \mbox{of degree less than
k}$, with any choice of the coefficients $a_l,\,l=0,\ldots,k$.
Therefore:
\begin{equation}\label{vdm1}
\prod_{i<j}^N(\lambda_i-\lambda_j)=\frac{(-1)^{\frac{N(N-1)}{2}}}
{a_0a_1...a_{N-1}}\det{
\left(\begin{array}{ccc}\pi_0(\lambda_1)&\ldots& \pi_0(\lambda_N)
\\ \pi_1(\lambda_1)&\ldots &\pi_1(\lambda_N)\\
.&.&.\\ .&.&.\\.&.&.\\ \pi_{N-1}(\lambda_1)&\ldots& \pi_{N-1}(\lambda_1)
\end{array}\right)}\equiv \frac{(-1)^{\frac{N(N-1)}{2}}}
{a_0a_1...a_{N-1}}\det{\left(
\pi_{i-1}(\lambda_j)\right)_{1\le i,j\le N}}.
\end{equation}
Multiplying every entry in $j_{th}$ column in the above determinant
with the factor $e^{-\frac{1}{2}Q(\lambda_j)}$ we
see that the JPDF can be conveniently written, up to a multiplicative
constant,  as
\begin{equation}\label{JPD1}
{\cal P}(\lambda_1,\ldots,\lambda_N)\propto
\left[\det{\left(e^{-\frac{1}{2}Q(\lambda_j)}\pi_{i-1}(\lambda_j)
\right)_{1\le i,j\le N}}\right]^2.
\end{equation}
If we let $\hat{A}$ be the matrix with the entries $A_{ij}=
\left(\phi_{i-1}(x_j)\right)_{1\le i,j\le N}$, then
\begin{eqnarray}\label{orp}
[\det{A}]^2=\det{\hat{A}^T\hat{A}}=
\det{\left(\sum_{j=1}^n A_{ji}A_{jk}\right)}.
\end{eqnarray}
This implies the following form of the JPDF:
\begin{equation}\label{JPD2}
{\cal P}(\lambda_1,\ldots,\lambda_N)\propto
\det{\left(\sum_{j=1}^N \phi_{j-1}(\lambda_i)\phi_{j-1}(\lambda_k)
\right)_{1\le i,k\le N}}\equiv
\det{\left(K_N(\lambda_i,\lambda_k)\right)_{1\le i,k\le N}}
\end{equation}
where we introduced the notation:
\begin{equation}\label{kern}
K_N(\lambda,\lambda')=\sum_{j=0}^{N-1}\phi_j(\lambda)\phi_j(\lambda')
\end{equation}
usually called ``kernel" in the literature. In our particular case
\begin{equation}\label{orp1}
\phi_{i-1}(\lambda)=
e^{-\frac{1}{2}Q(\lambda)}\pi_{i-1}(\lambda)
\end{equation}
so that the kernel is given explicitly by
\begin{equation}\label{kern1}
K_N(\lambda,\lambda')=e^{-\frac{1}{2}\left(Q(\lambda)+Q(\lambda')\right)}
\sum_{j=0}^{N-1}\pi_{j}(\lambda)\pi_{j}(\lambda').
\end{equation}
Now it is easy to see that if we take the polynomials $\pi_{i}(x)$
such that they form an {\it orthonormal system} with respect to
the weight $e^{-Q(x)}$, the corresponding kernel will be a
``reproducing" one with respect to the measure $d\mu(x)\equiv dx$,
in the sense of the ``integrating-out" Lemma. Indeed, suppose that
$\pi_i(x)$ satisfy the orthonormality conditions:
\begin{equation}\label{orp2}
\int e^{-Q(x)}\pi_i(x)\pi_j(x)\,dx=\delta_{ij},
\end{equation}
for any indices $i\ge 1,\,j\ge 1$.
Then we obviously have
\begin{eqnarray}\label{kern2}
\nonumber &&
\int K_N(x,y)K_N(y,z)dy=\sum_{j=0}^{N-1}\sum_{k=0}^{N-1}
e^{-\frac{1}{2}\left(Q(x)+Q(z)\right)}\pi_{j}(x)\pi_{k}(z)
\int \pi_{j}(y)\pi_{k}(y)e^{-Q(y)}dy\\
&&=\sum_{j=0}^{N-1}e^{-\frac{1}{2}\left(Q(x)+Q(z)\right)}
\pi_{j}(x)\pi_{j}(z)=K_N(x,z)
\end{eqnarray}
exactly as required by the reproducing property. Moreover, in this
case obviously
\[
q_N=\int K_N(x,x)dx=\sum_{j=0}^{N-1}\int\,
e^{-Q(x)}\pi_j(x)\pi_j(x)\,dx=N,
\]
and therefore the relation (\ref{32}) amounts to
\begin{equation}\label{kern3}
\int \mbox{det}\left(K_N(x_i,x_j)\right)_{1\le i,j\le N}
\,dx_N=\mbox{det}\left(K_N(x_i,x_j)\right)_{1\le i,j\le N-1}.
\end{equation}
Continuing this process one step further we see
\begin{eqnarray}\label{kern4}
&&\int\ \int \mbox{det}\left(K_N(x_i,x_j)\right)_{1\le i,\,j\le N}
\,dx_{N-1}dx_{N}=
\int \mbox{det}\left(K_N(x_i,x_j)\right)_{1\le i,\,j\le N-1}dx_{N-1}
\\
&=& [N-(N-2)]\mbox{det}\left(K_N(x_i,x_j)\right)_{1\le i,\,j\le N-2}
\end{eqnarray}
and continuing by induction
\begin{equation}\label{kern5}
\int\ldots \int\mbox{det}
\left(K_N(x_i,x_j)\right)_{1\le i,\,j\le N}
\,dx_{k+1}\ldots dx_{N}=
(N-k)!\,\mbox{det}\left(K_N(x_i,x_j)\right)_{1\le i,\,j\le k}
\end{equation}
for $k=1,2,\ldots$, and the result is $N!$ for $k=0$. Remembering
the expression of the JPDF, Eq.(\ref{JPD2}), in terms of the
kernel $K_N(x_i,x_j)$ we see that, in fact, the theory developed
provided simultaneously the explicit formulae for all $n-$point
correlation functions of the eigenvalues ${\cal R}_n(\lambda_1,
\ldots, \lambda_n)$, introduced by us earlier, Eq.(\ref{23}):
\begin{equation}\label{kern6}
{\cal R}_n(\lambda_1, \ldots, \lambda_n)=
\mbox{det}\left(K_N(\lambda_i,\lambda_j)\right)_{1\le i,j\le n}
\end{equation}
expressed, in view of the relations Eq.(\ref{kern1})
effectively in terms of the orthogonal polynomials $\pi_k(\lambda)$.
In particular, remembering the relation between the mean eigenvalue
density and the one-point function derived by us earlier, we have:
\begin{equation}\label{kern7}
\overline{\rho_N(\lambda)}=K_N(\lambda,\lambda)=
\sum_{j=1}^{N-1}
e^{-Q(\lambda)}\pi_{j-1}(\lambda)\pi_{j-1}(\lambda).
\end{equation}
The latter result allows to represent the ``connected" (or
``cluster") part of the two-point correlation function introduced
by us in Eq.(\ref{sigma2}) in the form:
\begin{equation}\label{kern8}
Y_2(\lambda_1,\lambda_2)=\overline{\rho_N(\lambda_1)}\,\,\,
\overline{\rho_N(\lambda_2)} -{\cal R}_2(\lambda_1, \lambda_2)
=\left[K_N(\lambda_1,\lambda_2)\right]^2.
\end{equation}

Finally, combining the relation Eq.(\ref{hole3}) between the hole probability
$A(L)$ and the n-point correlation functions, and on the other hand
the expression of the latter in terms of the kernel
$K_N(\lambda,\lambda')$, see Eq.(\ref{kern6}), we arrive at
\begin{eqnarray}\label{holekern}
 A(L)=\sum_{j=0}^N\frac{(-1)^j}{j!} \int_{-L/2}^{L/2}\ldots
\int_{-L/2}^{L/2}\, \det{\left(\begin{array}{ccc}
K_N(\lambda_1,\lambda_1)&\ldots&K_N(\lambda_1,\lambda_j)\\
.&.&.\\.&.&.\\.&.&.\\
K_N(\lambda_j,\lambda_1)&\ldots&K_N(\lambda_j,\lambda_j)
\end{array}\right)}
\,\, d\lambda_1 \ldots d\lambda_j.
\end{eqnarray}
In fact, the last expression can be written in a very compact form by
noticing that it is just a Fredholm determinant
$\det{\left({\cal I}-{\cal K}_N\right)}$, where ${\cal K}_N$
is a (finite rank) integral operator with the kernel
$K_N(\lambda,\lambda')=\sum_{i=0}^{N-1}\phi_i(\lambda)\phi_i(\lambda')$
acting on square-integrable functions on the interval $\lambda\in (-L/2,L/2)$.

\section{Properties of Hermite polynomials}
\subsection{Orthogonality, Recurrent Relations
and Integral Representation}

Consider the set of polynomials $h_k(x)$ defined as
\footnote{The standard reference to the Hermite polynomials
uses the definition
\[
H_k(x)=(-1)^ke^{x^2}
\frac{d^k}{dx^k}\left(e^{-x^2}\right)=2^kx^k+\cdots,
\]
Such a choice ensures $H_k(x)$ to be
orthogonal with respect to the weight $e^{-x^2}$. Our choice
is motivated by random matrix applications, and is related to the
standard one as $h_k(x)=H_k\left(\sqrt{\frac{N}{2}x}\right)$.
}
\begin{equation}\label{Her1}
h_k(x)=(-1)^ke^{N\frac{x^2}{2}}
\frac{d^k}{dx^k}\left(e^{-N\frac{x^2}{2}}\right)=N^kx^k+\cdots,
\end{equation}
and consider, for $k\ge l$
\begin{eqnarray}\label{Her2}
&&\int_{-\infty}^{\infty}
e^{-N\frac{x^2}{2}}h_l(x)h_k(x)dx=
(-1)^k\int_{-\infty}^{\infty}\,dx\,  h_l(x)
\frac{d^k}{dx^k}\left(e^{-N\frac{x^2}{2}}\right)\\ \nonumber
&&=(-1)^{k+1}\int_{-\infty}^{\infty}\,dx\,  h'_l(x)
\frac{d^{k-1}}{dx^{k-1}}\left(e^{-N\frac{x^2}{2}}\right)
=\ldots=(-1)^{2k}\int_{-\infty}^{\infty}\,dx\, \,e^{-N\frac{x^2}{2}}
\frac{d^k}{dx^{k}} h_l(x).
\end{eqnarray}
Obviously, for $k>l$ we have $\frac{d^{k}}{dx^{k}} h_l(x)=0$,
whereas for $k=l$ we have $\frac{d^{k}}{dx^{k}} h_k(x)=k!N^k$. In
this way we verified the orthogonality relations and the
normalization conditions
\begin{eqnarray}\label{Her3}
\int_{-\infty}^{\infty}
e^{-N\frac{x^2}{2}}\tilde{h}_l(x)\tilde{h}_k(x)dx= \delta_{kl}
\end{eqnarray}
for normalized polynomials
\begin{equation}\label{normher}
\tilde{h}_k(x)=:\frac{1}{\left[k!N^k\sqrt{\frac{2\pi}{N}}\right]^{1/2}}
h_k(x).
\end{equation}
In the theory of orthogonal polynomials an important role is played
by recurrence relations:
\begin{eqnarray}\label{Her4}& h_{k+1}(x)=(-1)^{k+1} e^{N\frac{x^2}{2}}
\frac{d^k}{dx^k} \left(\frac{d}{dx}e^{-N\frac{x^2}{2}}\right)=
(-1)^{k+2}N e^{N\frac{x^2}{2}}\frac{d^k}{dx^k}
\left(xe^{-N\frac{x^2}{2}}\right)\\ &
 \nonumber =(-1)^{k+2}N e^{N\frac{x^2}{2}}
\left[\left(\begin{array}{c}0\\ k\end{array}
\right)x\frac{d^k}{dx^k}\left(e^{-N\frac{x^2}{2}}\right)+
\left(\begin{array}{c}1\\ k\end{array}
\right)\frac{d^{k-1}}{dx^{k-1}}\left(e^{-N\frac{x^2}{2}}\right)\right]
=N\left[x\,h_k(x)-k\,h_{k-1}(x)\right],
\end{eqnarray}
where we exploited the Leibniz formula for the $k-$th derivative
of a product. After normalization we therefore have
\begin{equation}\label{Her5}
\left[\frac{k+1}{N}\right]^{1/2}\tilde{h}_{k+1}(x)=x\,
\tilde{h}_{k}(x)-\left[\frac{k}{N}\right]^{1/2}\tilde{h}_{k-1}(x).
\end{equation}
Let us multiply this relation with $\tilde{h}_k(y)$, and then
replace $x$ by $y$. In this way we arrive at two relations:
\begin{eqnarray}\label{Her6}
&& \left[\frac{k+1}{N}\right]^{1/2}\tilde{h}_{k+1}(x)
\tilde{h}_{k}(y)=x\,
\tilde{h}_{k}(x)\tilde{h}_{k}(y)-
\left[\frac{k}{N}\right]^{1/2}\tilde{h}_{k-1}(x)\tilde{h}_{k}(y),\\
&& \left[\frac{k+1}{N}\right]^{1/2}\tilde{h}_{k+1}(y)
\tilde{h}_{k}(x)=y\,
\tilde{h}_{k}(x)\tilde{h}_{k}(y)-
\left[\frac{k}{N}\right]^{1/2}\tilde{h}_{k-1}(y)\tilde{h}_{k}(x).
\end{eqnarray}
The difference between the upper and the lower line can be
written for any $k=1,2,\ldots $ as
\[
(x-y)\tilde{h}_{k}(x)\tilde{h}_{k}(y)=A_{k+1}-A_k\,,\quad A_k=
\left[\frac{k}{N}\right]^{1/2}\{\tilde{h}_{k-1}(y)\tilde{h}_{k}(x)-
\tilde{h}_{k-1}(x)\tilde{h}_{k}(y)\}.
\]
Summing up these expressions over $k$:
\[
(x-y)\sum_{k=1}^{n-1}\tilde{h}_{k}(x)\tilde{h}_{k}(y)=(A_2+\ldots+A_{n})-
(A_1+\ldots+A_{n-1})=A_{n}-A_1
\]
and remembering that
$A_1=\sqrt{\frac{N}{2\pi}}(x-y)=(x-y)\tilde{h}_0(x)\tilde{h}_0(y)$
we arrive at a very important relation:
\begin{equation}\label{darboux}
\sum_{k=0}^{n-1}\tilde{h}_{k}(x)\tilde{h}_{k}(y)=
\sqrt{\frac{n}{N}}\frac{\tilde{h}_{n-1}(y)\tilde{h}_{n}(x)-
\tilde{h}_{n-1}(x)\tilde{h}_{n}(y)}{x-y},
\end{equation}
or, for the original (not-normalized) polynomials:
\begin{equation}\label{darboux1}
\sum_{k=0}^{n-1}\frac{1}{k!N^k}h_{k}(x)h_{k}(y)=\frac{1}{(n-1)!N^{n}}
\frac{h_{n-1}(y)h_{n}(x)-h_{n-1}(x)h_{n}(y)}{x-y},
\end{equation}
which are known as the {\it Christoffel-Darboux formulae}.
Finally, taking the limit $x\to y$ in the above expression we see
that
\begin{equation}\label{darboux2}
\sum_{k=0}^{n-1}\frac{1}{k!N^k}h^2_{k}(x)=\frac{1}{(n-1)!N^{n}}
\left[h_{n-1}(x)h'_{n}(x)-h'_{n-1}(x)h_{n}(x)\right].
\end{equation}

Most of the properties and relations discussed above for Hermite
polynomials have their analogues for general class of orthogonal
polynomials. Now we are going to discuss another very useful
property which is however shared only by few families of {\it
classical} orthogonal polynomials: Hermite, Laguerre, Legendre,
Gegenbauer and Jacoby. All these polynomials have one of few {\it
integral representations} which are frequently exploited when
analyzing their properties. For the case of Hermite polynomials we
can most easily arrive to the corresponding representation by
using the familiar Gaussian integral identity, cf.
Eq.(\ref{Gauint0}):
\begin{equation}\label{Gauint}
e^{-N\frac{x^2}{2}}=\sqrt{\frac{N}{2\pi}}\int_{-\infty}^{\infty}
\, dq\, e^{-\frac{N}{2}q^2+ixqN}.
\end{equation}
Substituting such an identity to the original definition,
Eq.(\ref{Her1}), we immediately see that
\begin{equation}\label{intrep}
h_k(x)=(-iN)^k\sqrt{\frac{N}{2\pi}}\,
e^{N\frac{x^2}{2}}\int_{-\infty}^{\infty}
\, dq\, q^k\,e^{-\frac{N}{2}q^2+ixqN},
\end{equation}
which is the required integral representation, to be mainly used
later on when addressing the large-$N$ asymptotics of the Hermite
polynomials. Meanwhile, let us note that differentiating the above
formula with respect to $x$ one arrives at the useful relation
$\frac{d}{dx}h_k(x)=Nxh_k(x)-h_{k+1}(x)=Nk\,h_{k-1}(x)$. This can
be further used to simplify the formula Eq.(\ref{darboux2})
bringing it to the form
\begin{equation}\label{darboux3}
\sum_{k=0}^{n-2}\frac{1}{k!N^k}h^2_{k}(x)=\frac{1}{(n-2)!N^{n-1}}
\left[h^2_{n}(x)-h_{n-1}(x)h_{n+1}(x)\right].
\end{equation}

\subsection{Saddle-point method and Plancherel-Rotach asymptotics
of Hermite polynomials} In our definition, the Hermite polynomials
$h_k(x)$ depend on two parameters: explicitly on the order index
$k=0,1,\ldots$ and implicitly on the parameter $N$ due to the fact
that the weight function $e^{-N\frac{x^2}{2}}$ contains this
parameter. Invoking the random matrix background for the use of
orthogonal polynomials, we associate the parameter $N$ with the
size of the underlying random matrix. From this point of view, the
limit $N\gg 1$ arises naturally as we are interested in
investigating the spectral characteristics of large matrices. A
more detailed consideration reveals that, from the random matrix
point of view, the most interesting task is to extract the
asymptotic behaviour of the Hermite polynomials with index $k$
large and comparable with $N$, i.e. $k=N+n$, where the parameter
$n$ is considered to be of the order of unity. Such behaviour is
known as Plancherel-Rotach asymptotics.

To understand this fact it is enough to invoke the relation
(\ref{kern7}) expressing the mean eigenvalue density in terms of
the set of orthogonal polynomials:
\begin{eqnarray}\label{kern77}
&&\overline{\rho_N(\lambda)}=K_N(\lambda,\lambda)=
e^{-\frac{N}{2}\lambda^2}\,\sum_{j=0}^{N-1}
\tilde{h}^2_{j}(\lambda),
\\
&& =e^{-\frac{N}{2}\lambda^2}\frac{\sqrt{N/2\pi}}{(N-1)!N^{N}}
\left[h^2_{N}(\lambda)-h_{N-1}(\lambda)h_{N+1}(\lambda)\right],
\label{kern777}
\end{eqnarray}
where we used the expressions pertinent to the Gaussian weight:
$Q(\lambda)\equiv\frac{N}{2}\lambda^2\,,\, \pi_k(\lambda)\equiv
\tilde{h}_k(\lambda)$, and further exploited the variant of the
Christoffel-Darboux formula, Eq.(\ref{darboux3}). It is therefore
evident that the limiting shape of the mean eigenvalue density for
large random matrices taken from the Gaussian Unitary Ensemble is
indeed controlled by the Plancherel-Rotach asymptotics of the
Hermite polynomials. In fact, similar considerations exploiting
the original Christoffel-Darboux formula, Eq.(\ref{darboux}), show
that our main object of interest -the kernel
$K_N(\lambda,\lambda')$ - can be expressed as
\begin{equation}\label{kernher}
K_N(\lambda,\lambda')=e^{-\frac{N}{4}(\lambda^2+\lambda'^2)}
\frac{\tilde{h}_{N-1}(\lambda)\tilde{h}_{N}(\lambda')-
\tilde{h}_{N-1}(\lambda)\tilde{h}_{N}(\lambda')}{\lambda-\lambda'}
\end{equation}
and therefore all the higher correlation functions are controlled
by the Plancherel-Rotach asymptotics as well.

For extracting the required asymptotics we are going to use the
integral representation for the Hermite polynomials. We start with
rewriting the expression Eq.(\ref{intrep}) as
\begin{eqnarray}\label{intrep1}
&& h_{N+n}(x)=(-iN)^{N+n}\sqrt{\frac{N}{2\pi}}\,
\int_{-\infty}^{\infty}
\, dq\, q^{N+n}\, e^{-\frac{N}{2}\left(q-ix\right)^2}
\\&=&(-iN)^{N+n}\sqrt{\frac{N}{2\pi}}\,\left[I_{N+n}(x)+
(-1)^{N+n}I_{N+n}(-x) \right],
\end{eqnarray}
where
\begin{equation}\label{integral}
I_{N+n}(x)=\int_{0}^{\infty}
\, dq\, q^{n}\, e^{Nf(q)},\quad
f(q)=\ln{q}-\frac{1}{2}\left(q-ix\right)^2.
\end{equation}
The latter form is suggestive of exploiting the so-called {\it
saddle-point} method (also known as the method of {\it steepest
descent} or method of {\it stationary phase}) of asymptotic
evaluation of integrals of the form
\begin{equation}\label{sp1}
\int_{\Gamma}\phi(z)e^{NF(z)}dz,
\end{equation}
where the integration goes along a contour $\Gamma$ in the complex
plane, $F(z)$ is an analytic function of $z$ in some domain
containing the contour of integration, and $N$ is a large
parameter. The main idea of the method can be informally outlined
as follows. Suppose that the contour $\Gamma$ is such that: (i)
the value of $\mbox{Re}F$ has its {\it maximum} at a point $z_0\in
\Gamma$, and decreases fast enough when we go along $\Gamma$ away
from $z_0$, and (ii) the value of $\mbox{Im}F$ stays constant
along $\Gamma$ (to avoid fast oscillations of the integrand). Then
we can expect the main contribution for $N\gg 1$ to come from a
small vicinity of $z_0=x_0+iy_0$.

\begin{figure}[h!]
\centering \epsfig{file=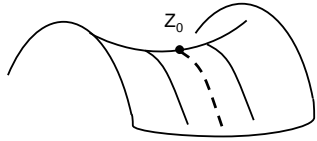, width=4.5cm} \caption{Schematic
structure of a harmonic function in the vicinity of a stationary
point $z_0$.} \label{fig2}\end{figure}

Since the function $\mbox{Re}F$ is a harmonic function of
$x=\mbox{Re}z,\, y=\mbox{Im}z$, it can have only {\it saddle
points} (see Fig.~\ref{fig2}) found from the condition of
stationarity $F'(z_0)=0$. Let us suppose that there exists only
one such saddle point $z=z_0$, close to which we can expand
$F(z)\approx F(z_0)+C(z-z_0)^2$, where $C=\frac{1}{2}F''(z_0)$.
Consider the level curves $[\mbox{Re}F](x,y)
=[\mbox{Re}F](x_0,y_0)$, which are known either to go to infinity,
or end up at a boundary of the domain of analyticity. In the
vicinity of the chosen saddle-point the equation for the level
curves is $\mbox{Re}[F(z)-F(z_0)]=0$, hence
\[
\mbox{Re}[|C|e^{i\theta}(z-z_0)^2]=\left[(x-x_0)^2
(y-y_0)^2\right]\cos{\theta}-2\,(x-x_0)(y-y_0)\sin{(\theta)}=0,
\]
which describes two orthogonal straight lines passing through the
saddle-point
\[
y=y_0+\tan{\left(\frac{\pi}{4}-\frac{\theta}{2}\right)}(x-x_0),\quad
y=y_0-\tan{\left(\frac{\pi}{4}+\frac{\theta}{2}\right)}(x-x_0)
\]
\begin{figure}[h!]
\centering\epsfig{file=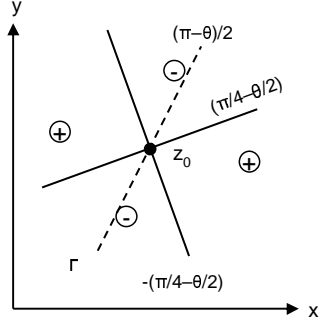, width=4.5cm}
\caption{Partitioning of the $x-y$ plane in a vicinity of the
stationary point $z_0$ into four sectors by ``level curves" (solid
lines). Dashed line shows the bi-sector of the negative sectors:
the direction of the steepest descent contour.} \label{fig3}
\end{figure}
partitioning the $x,y$ plane into four sectors: two ``positive"
ones: $\mbox{Re}F(z)>\mbox{Re}F(z_0)$, and two ``negative" ones
$\mbox{Re}F(z)<\mbox{Re}F(z_0)$, see Fig.~\ref{fig3}. If the
``edge points" of the integration contour $\Gamma$ (denoted $z_1$
and $z_2$) both belong to the {\it same} sector, and
$\mbox{Re}F(z_1)\ne \mbox{Re}F(z_2)$, one always can deform the
contour in such a way that $\mbox{Re}F(z)$ is monotonically
increasing along the contour. Then obviously the main contribution
to the integral comes from the vicinity of the endpoint (of the
largest value of $\mbox{Re}F(z)$). Essentially the same situation
happens when $z_1$ belongs to a negative (positive) sector, and
$z_2$ is in a positive (resp., negative) sector. And only if the
two endpoints belong to two {\it different negative} sectors, we
can deform the contour in such a way, that $\mbox{Re}F(z)$ has its
maximum along the contour at $z=z_0$, and decays away from this
point. Moreover, it is easy to understand that the fastest decay
away from $z_0$ will occur along the {\it bi-sector} of the
negative sectors, i.e. along the line
$y-y_0=\tan{\frac{\pi-\theta}{2}}(x-x_0)$. Approximating the
integration contour in the vicinity of $z_0$ as this bi-sector,
i.e. by
$z=z_0+(x-x_0)\frac{e^{-i\frac{\pi-\theta}{2}}}{\sin{(\theta/2)}}$,
we get the leading term of the large-$N$ asymptotics for the
original integral by extending the limits of integration in the
variable $\tilde{x}=x-x_0$ from $-\infty$ to $\infty$:
\begin{eqnarray}\label{sp2}
&&\nonumber \int_{\Gamma}\phi(z)e^{NF(z)}dz\approx \phi(z_0)e^{NF(z_0)}
\frac{e^{-i\frac{\pi-\theta}{2}}}{\sin{(\theta/2)}}
\int_{-\infty}^{\infty}d\tilde{x}e^{-N|C|\tilde{x}^2}{\sin^2{\theta/2}}\\
&&=\phi(z_0)\sqrt{\frac{2\pi}{N|F''(z_0)|}}
\exp{\{NF(z_0)+\frac{i}{2}(\pi-Arg[F''(z_0)/2)])\}}.
\end{eqnarray}
It is not difficult to make our informal consideration rigorous,
and to calculate systematic corrections to the leading-order result,
as well as to consider the case of several isolated saddle-points,
the case of a saddle-point coinciding with an end of the contour, etc.,
see \cite{saddlepoint} for more detail.

After this long exposition of the method we proceed by applying it
to our integral, Eq.(\ref{integral}). The saddle-point equation
and its solution in that case amount to:
\[
F'(q)=\frac{1}{q}-q+ix=0,\quad q=q_{\pm}=\frac{1}{2}\left(
ix\pm \sqrt{4-x^2}\right).
\]
It is immediately clear that we have essentially three different
cases: a) $|x|<2$ (b) $|x|>2$ and (c) $|x|=2$.
\begin{enumerate}
\item {\bf $|x|<2$}. In this case we can introduce
$x=2\cos{\phi},\,\, 0<\phi<\pi$, so that $q_{\pm}=i\cos{\phi}\pm
\sin{\phi}$, or $q_{+}=e^{-i(\phi-\pi/2)},
\,q_{-}=e^{i(\phi+\pi/2)}$. It is easy to understand that we are
interested only in $q_{+}$ (see Fig.4) and to calculate that
$\mbox{Re}f(q_{+})=\frac{1}{2}\cos{(2\phi)}$. On the other hand
$\mbox{Re}f(q)\to-\infty$ when either $q\to \infty$ or $q\to 0$,
so that both endpoints belong to negative sectors. To understand
whether they belong to the same or different sectors, we consider
the values of $\mbox{Re}f(q)=\ln{R}-\frac{1}{2}(R^2-x^2)$ along
the real axis, $q=R$-real. As a function of the variable $R$ this
expression has its maximal value
$\mbox{Re}f(q=1)=-\frac{1}{2}+2\cos^2{\phi}$ at $q=R=1$.

\begin{figure}[h!]
\centering\epsfig{file=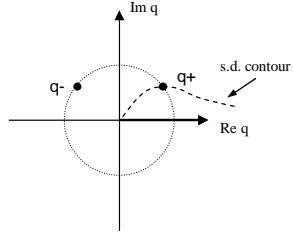, width=4.0cm} \caption{Structure
of the saddle-points $q_{\pm}$ and the relevant steepest descent
contour for $|x|<2$.} \label{fig4}\end{figure}

Noting that $\mbox{Re}f(q=1)-\mbox{Re}f(q_{+})=\cos^2{\phi}>0$, we
conclude that the point $q=1$ belongs to a positive sector, and
therefore the existence of this positive sector makes the
endpoints $q=0$ and $q=\infty$ belonging to two {\it different}
negative sectors, as required by the saddle-point method.
Calculating
\[
f''(q_{+})=-\left(1+\frac{1}{q^2_{+}}\right)=2i\sin{\phi}e^{i\phi}
\]
we see that $|C|=\sin{\phi},\, \theta=\phi+\pi/2$, and further
\[
f''(q_{+})=\frac{1}{2}\cos{(2\phi)}+i\left[\frac{1}{2}\sin{(2\phi)}-
\phi+\pi/2\right].
\]
Now we have all the ingredients to enter in Eq.(\ref{sp2}), and
can find the leading order contribution to $I_{N+n}(x)$. Further
using $I_{N+n}(-x)=\overline{I_{N+n}(x)}$, valid for real $x$, we
obtain the required Plancherel-Rotach asymptotics of the Hermite
polynomial:
\begin{eqnarray}\label{planrot1}
h_{N+n}(x)\approx N^{N+n}\sqrt{\frac{2}{\sin{\phi}}}
e^{\frac{N}{2}\cos{2\phi}}\cos{\left\{(n+1/2)\phi-\pi/4+N
\left(\phi-\frac{1}{2}\sin{2\phi}\right)\right\}},
\end{eqnarray}
where $x=2\cos{\phi},\quad 0<\phi<\pi,\quad n\ll N$.

Now we consider the opposite case:
\item $|x|>2$. It is enough to consider explicitly
the case $x>2$ and parameterize $x=2\cosh{\phi},\quad
0<\phi<\infty$. The saddle points in this case are purely
imaginary:
\begin{equation}\label{out1}
q_{\pm}=\frac{i}{2}(2\cosh{\phi}\pm 2\sinh{\phi})=ie^{\pm \phi}.
\end{equation}
One possible contour of the constant phase passing through both
points is just the imaginary axis $q=iy$, where
$\mbox{Im}f(q)=\pi/2$ and
$\mbox{Re}f(q)=\ln{y}+\frac{1}{2}(y-x)^2.$ Simple consideration
gives that $y_{-}=e^{-\phi}$ corresponds to the maximum, and
$y_{+}=e^{\phi}$ to the minimum of $\mbox{Re}f(q)$ along such a
contour. It is also clear that for $q=iy_{+}$ the expression
$\mbox{Re}f(q)$ has a local maximum along the path going through
this point in the direction {\it transverse} to the imaginary
axis. The ``topography" of $\mbox{Re}f(q)$ in the vicinity of the
two saddle-points  is sketched in Fig.~\ref{fig5}

\begin{figure}[h!]\centering\epsfig{file=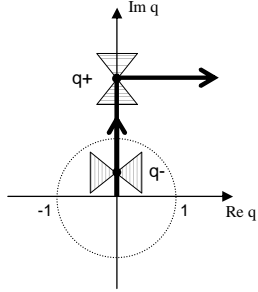,width=3.5cm}
\caption{The saddle-points $q_{\pm}$, the corresponding positive
sectors (shaded), and the relevant steepest descent contour (bold)
for $|x|>2$.}\label{fig5}\end{figure}

This discussion suggests a possibility to deform the path of
integration $\Gamma$ to be a contour of constant phase
$\mbox{Im}f(q)$ consisting of two pieces - $\Gamma_1=\{q=iy,\,0\le
y\le y_{+}\}$ and $\Gamma_2$ starting from $q=iy_{+}$
perpendicular to the imaginary axis and then going towards
$q=\infty$. Correspondingly,
\begin{equation}\label{offspec}
I_{N+n}(x>2)=\int_{0}^{\infty}\, dq\, q^{n}\, e^{Nf(q)}
=\int_{0}^{iy_{+}}
\, dq\, q^{n}\, e^{Nf(q)}+\int_{\Gamma_2}
\, dq\, q^{n}\, e^{Nf(q)}.
\end{equation}
The second integral is dominated by the vicinity of the
saddle-point $q=iy_{+}$, and its evaluation by the saddle-point
technique gives:
\[
\int_{\Gamma_2} \, dq\, q^{n}\, e^{Nf(q)}\approx
\frac{1}{2}\sqrt{\frac{\pi e^{\phi}}{N\sinh{\phi}}} i^{n+N}
e^{n\phi+N\left(\phi+\frac{1}{2}e^{-2\phi}\right)},
\]
where the factor $\frac{1}{2}$ arises due to the saddle-point
being simultaneously the end-point of the contour. As to the first
integral, it is dominated by the vicinity of $iy_{-}$, and can
also be evaluated by the saddle-point method. However, it is easy
to verify that when calculating $h_{N+n}(x)\propto
\left[I_{N+n}(x)+ (-1)^{N+n}I_{N+n}(-x) \right] $ the
corresponding contribution is cancelled out. As a result, we
recover the asymptotic behaviour of Hermite polynomials for $x>2$
to be given by:
\begin{equation}\label{offspec1}
h_{N+n}(x=2\cosh{(2\phi)}>2)=\frac{N^{n+N}
e^{-\frac{N}{2}}}{\sqrt{\sinh{\phi}}}
e^{\left(n+\frac{1}{2}\right)\phi-
\frac{N}{2}\left(\sinh{(2\phi)}-2\phi\right)}.
\end{equation}

Now we come to the only remaining possibility, \item $|x|=2$. It
is again enough to consider only the case $x=2$ explicitly. In
fact, this is quite a special case, since for $x\to 2$ two
saddle-points $q_{\pm}$ degenerate into one: $q_{+}\to q_{-}\to
i$. Under such exceptional circumstances the standard saddle-point
method obviously fails. Indeed, the method assumed that different
saddle-points do not interfere, which means the distance
$|q_{+}-q_{-}|=\sqrt{|4-x^2|}$ is much larger than the typical
widths $W\sim \frac{1}{\sqrt{N|f''(q_{\pm})|}}$ of the regions
around individual saddle-points which yield the main contribution
to the integrand. Simple calculation gives
$|f''(q_{\pm})|=|1+q^{-2}_{\pm}|=\sqrt{|4-x^2|}$, and the
criterion of two separate saddle-points amounts to $|x-2|\gg
N^{-2/3}$. We therefore see that in the vicinity of $x=2$ such
that $|x-2|\sim N^{-2/3}$ additional care must be taken when
extracting the leading order behaviour of the corresponding
integral $I_{N+n}(x=2)$ as $N\to\infty$.

To perform the corresponding calculation, we introduce a new
scaling variable $\xi=N^{2/3}(2-x)$, and consider $\xi$ to be
fixed and finite when $N\to\infty$. We also envisage from the
discussion above  that the main contribution to the integral comes
from the domain around the saddle-point $q_{sp}=i$ of the widths
$|q-i|\sim \sqrt{|2-x|}\sim N^{-1/3}$. The integral we are
interested in is given by
\begin{eqnarray}\label{ai1}
J_N(\xi)&=&\int_{-\infty}^{\infty}
\, dq\, q^{N+n}\, e^{-\frac{N}{2}\left(q-ix\right)^2}\\
&=&N^{-1/3} \int_{-\infty}^{\infty}
\, dt\, \left(i+\frac{t}{N^{1/3}}\right)^n\,
e^{N\left[\ln{(i+\frac{t}{N^{1/3}})}-
\frac{1}{2}\left\{i+\frac{t}{N^{1/3}}-i\left(2-\frac{\xi}{N^{2/3}}
\right)\right\}^2\right]}
\end{eqnarray}
where we shifted the contour of integration from the real axis to
the line $q=i+\frac{t}{N^{1/3}}\,, -\infty<t<\infty$ to ensure
that it passes through the expected saddle-point $q_{sp}=i$, and
also scaled the integration variable appropriately. Now we can
consider $\xi,t$-finite when $N\gg 1$, and expand the integrand
accordingly. A simple computation yields:
\begin{eqnarray}\label{ai2}
J_{N\gg 1}(\xi)\approx
N^{-1/3}i^{N+n}\,e^{N/2-N^{1/3}\xi} \int_{-\infty}^{\infty}
\, dt\, e^{-i\xi\,t+i\frac{t^3}{3}}.
\end{eqnarray}
Up to a constant factor the integral appearing in this expression
is, in fact, a representation of a special function known as Airy
function $Ai(\xi)$:
\begin{equation}\label{Airydef}
Ai(\xi)=\frac{1}{\pi}\int_{0}^{\infty}
\, dt\,\, \cos{\left(\xi\,t+\frac{t^3}{3}\right)}
\end{equation} which is a solution
of the second-order linear differential equation
$\frac{d^2}{d\xi^2}F(\xi)-\xi\,F(\xi)=0$. A typical behaviour of
such a solution is shown in Fig.~\ref{fig6}.

\begin{figure}[h!]
\centering\epsfig{file=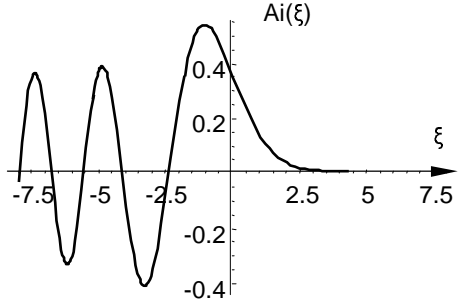} \caption{The Airy function
$Ai(\xi)$.} \label{fig6}\end{figure}

All this results in the asymptotic behaviour of the Hermitian
polynomials in the so-called ``scaling vicinity" of the point
$x=2$:
\begin{eqnarray}\label{Air3}
h_{N+n}\left(x=2-\frac{\xi}{N^{2/3}}\right)\approx
\frac{N^{1/6}}{\sqrt{2\pi}}N^{N+n}\,e^{N/2-N^{1/3}\xi}\,Ai(-\xi).
\end{eqnarray}
Such scaling vicinity of $x=2$ is what gives room for a
transitional regime between the oscillating asymptotics of the
Hermite polynomials for $|x|<2$, see Eq.(\ref{planrot1}), and the
exponential decay typical for $|x|>2$ as described in
Eq.(\ref{offspec1}). Formula (\ref{Air3}) indeed matches
Eq.(\ref{planrot1}) as $\xi\to\infty$ and Eq.(\ref{offspec1}) as
$\xi\to-\infty$. This statement is most easily verified by
invoking the known asymptotics of the Airy function:
\begin{equation}\label{Airyas}
Ai(-\xi)\approx \left\{\begin{array}{c}\xi^{-1/4}\pi^{-1/2}
\cos{\left(-\frac{2}{3}\xi^{3/2}+\frac{\pi}{4}\right)},\quad \xi\to \infty\\
\frac{1}{\pi^{1/2}|\xi|^{1/4}}\,e^{-\frac{2}{3}|\xi|^{3/2}},\quad
\xi\to-\infty\end{array}\right.
\end{equation}
and identifying $\phi=|\xi|^{1/2}N^{-1/3}\ll 1$
in the corresponding expressions.
\end{enumerate}

Now we are going to apply the derived formulae for extracting the
large-N behaviour of the mean eigenvalue density and the kernel
 as described in Eqs.(\ref{kern77}) and(\ref{kernher}), respectively.
In fact, it is more conventional in the random matrix literature
to use the mean density to be normalized to unity, rather than to
$N$. Such a density will have a well-defined large-N limit which
we will denote as $\rho_{\infty}(\lambda)$.

\section{Scaling regimes for GUE}
\subsection{Bulk scaling: Wigner semicircle and Dyson kernel.}

\begin{figure}[h!]
\centering\epsfig{file=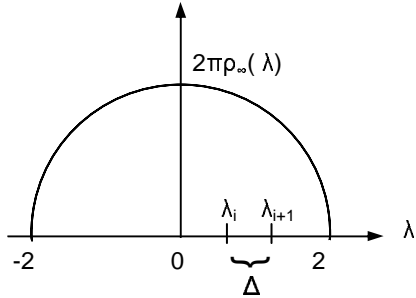} \caption{Wigner semicircular
density. Sketch shows a typical spacing between neighbouring
levels in the bulk of the spectrum.} \label{fig7}\end{figure}

The first case to be considered is the spectral parameter
$|\lambda|<2$ when we can parameterize $\lambda=2\cos{\phi}$, and
exploit the Plancherel-Rotach expression (\ref{planrot1}) for the
Hermite polynomials. Furthermore, denoting
$\alpha=\frac{1}{2}\phi-\frac{\pi}{4}+N
\left(\phi-\frac{1}{2}\sin{2\phi}\right)$, and using the identity
$\cos^2{\alpha}-\cos{(\alpha+\phi)}\cos{(\alpha-\phi)}=\sin^2{\phi}$
we find that
$h_N^2(\lambda)-h_{N-1}(\lambda)h_{N+1}(\lambda)\approx
2N^{2N}\sin{\phi} e^{N\cos{(2\phi)}}$. Furthermore, using for
large $N$ the Stirling formula: $(N-1)!\approx
\sqrt{\frac{2\pi}{N}}N^Ne^{-N}$ and remembering that
$\sin{\phi}=\frac{1}{2}\sqrt{4-\lambda^2}$ we arrive, after
collecting all factors, to the famous Wigner semicircular law for
the mean (normalized) spectral density:
\begin{equation}\label{semi}
\lim_{N\to\infty}\left[
\frac{1}{N}\overline{\rho(\lambda)}\right]=\rho_{\infty}(\lambda)=
\frac{1}{2\pi}\sqrt{4-\lambda^2},
\quad |\lambda|<2.
\end{equation}
We see that in the  limit of large $N$ all $N$ eigenvalues of GUE
matrices are concentrated in the interval $[-2,2]$, and the
typical separation of two neighbouring eigenvalues close to an
``internal" point $\lambda \in (-2,2)$ is
$\Delta=\frac{1}{N\rho_{\infty}(\lambda)}=O(N^{-1})$, see
Fig.~\ref{fig7}. That is why the case $\lambda \in (-2,2)$ is
frequently referred to as the "bulk of the spectrum" regime.

Let us now follow the same strategy for obtaining, under the same
conditions, the limiting expression for the kernel
$K(\lambda,\lambda')$ using for this goal formula (\ref{kernher}).
We have:
\begin{eqnarray}\label{kernher11}
&& h_N(\lambda)h_{N-1}(\lambda')-h_N(\lambda')h_{N-1}(\lambda)
\\ &&\approx 2N^{2N}\frac{1}{\sqrt{\sin{\phi}\sin{\phi'}}}
e^{\frac{N}{2}\left(\cos{(2\phi)}+\cos{(2\phi')}\right)}
\left[\cos{\alpha_1^{+}}\cos{\alpha_2^{-}}
-\cos{\alpha_1^{-}}\cos{\alpha_2^{+}}\right]
\end{eqnarray}
where $\alpha^{\pm}_{1}=\pm\frac{1}{2}\phi-\frac{\pi}{4}+N
\left(\phi-\frac{1}{2}\sin{2\phi}\right),\,\,
\alpha^{\pm}_{2}=\pm\frac{1}{2}\phi'-\frac{\pi}{4}+N
\left(\phi'-\frac{1}{2}\sin{2\phi'}\right)$. The next step is to
introduce $\psi=(\phi+\phi')/2$ and $\Omega= (\phi-\phi')/2$, and
to consider the parameter $\Omega$ to be of the order of
$O(N^{-1})$ when taking the limit. This choice ensures that the
distance $\lambda-\lambda'= 2[\cos{\phi}-\cos{\phi'}]\approx
4\Omega \sin{\psi}\approx 4\Omega \pi\rho_{\infty}(\lambda)$ is of
the order of the mean eigenvalue separation $\Delta$ - the typical
scale for the correlations between the eigenvalues in the {\it
bulk} of the spectrum- and thus must be reflected in the structure
of the kernel. To this end we denote $\Omega=\omega/N$, and keep
in the expressions for $\alpha^{\pm}_{1,2}$ terms up to the order
$O(1)$, i.e. writing $\alpha^{\pm}_{1,2}=N\beta+\left[
\pm\frac{1}{2}\psi-\frac{\pi}{4}\pm 2\omega \sin^2{\psi}\right]$,
where $\beta=\left(\psi-\frac{1}{2}\sin{2\psi}\right)$. With the
same precision:
\[
\cos{\alpha_1^{+}}\cos{\alpha_2^{-}}
-\cos{\alpha_1^{-}}\cos{\alpha_2^{+}}\approx \sin{\psi}
\sin{\left(4\omega\sin^2{\psi}\right)}\approx \sin{\psi}
\sin{\left[\pi \rho_{\infty}(\lambda)N(\lambda_1-\lambda_2)\right]}
\]
and $\cos{2\phi_1}+\cos{2\phi_2}\approx
2\left(\frac{\lambda^2}{2}-1\right)$ substituting all these
factors back into Eq.(\ref{kernher11}), we get
\begin{eqnarray}\label{kernher12}
h_N(\lambda)h_{N-1}(\lambda')- h_N(\lambda')h_{N-1}(\lambda)
\approx 2N^{2N} e^{2N\left(\frac{\lambda^2}{2}-1\right)}
\sin{\left[\pi \rho_{\infty}(\lambda)N(\lambda-\lambda')\right]}.
\end{eqnarray}
Now, taking into account the normalization factors in
$\tilde{h}_{N}(\lambda)$ and $\tilde{h}_{N-1}(\lambda),$  see
Eq.(\ref{normher}), using again the Stirling formula and invoking
Eq.(\ref{kern7}) we arrive at the following asymptotic expression
for the kernel, Eq.(\ref{kernher}):
\begin{equation}\label{Dyson}
\lim_{N\to\infty} \left[\frac{K_N(\lambda,\lambda')}
{K_N(\lambda,\lambda)}\right]=K_{\infty}\left[
N\rho_{\infty}(\lambda)(\lambda-\lambda')\right],\quad
K_{\infty}(r)=\frac{\sin{\pi r}}{\pi r}
\end{equation}
where $K_{\infty}(r)$ is the famous {\it Dyson scaling} form for
the kernel. The formula is valid as long as both $\lambda$ and
$\lambda'$ are within the range $(-2,2)$, and
$\lambda-\lambda'=O(N^{-1})$. Such choice of the parameters is
frequently referred to as the ``bulk scaling" limit.

\begin{figure}[h!]
\centering\epsfig{file=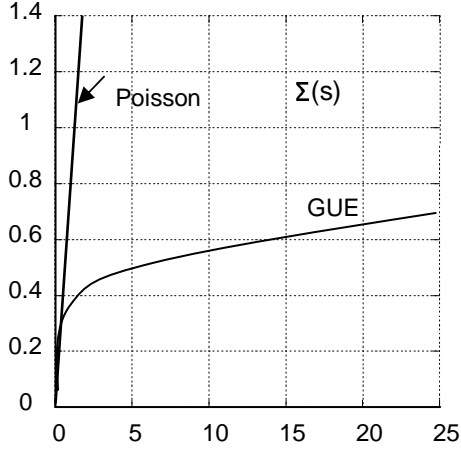} \caption{The logarithmic growth
of the number variance $\Sigma(s)$ for large GUE matrices versus
linear growth for uncorrelated (Poissonian) spectrum.}
\label{fig8}\end{figure}

Having at our disposal the limiting form of both mean eigenvalue
density and the two-point kernel we can analyse such important
statistical characteristics of the spectra as e.g. the ``number
variance", see Eq.(\ref{sigma2}), for an interval of the length
$L$ comparable with the mean spacing close to the origin
$\Delta=[N\rho_{\infty}(0)]^{-1}$. Under such a condition we can
legitimately employ the scaling form Eq.(\ref{Dyson}) of the
kernel when substituting it into formula (\ref{kern8}) for the
cluster function $Y_2(\lambda,\lambda')$. In this way we arrive at
\begin{eqnarray}\label{sigmasc}
\nonumber && \Sigma_2(L)=N\int_{-L/2}^{L/2} \rho_{\infty}(\lambda)d\lambda
-N^2 \int_{-L/2}^{L/2}d\lambda \, \int_{-L/2}^{L/2} d\lambda'
\rho_{\infty}(\lambda)\rho_{\infty}(\lambda')
K^2_{\infty}\left[
N\rho_{\infty}(\lambda)(\lambda-\lambda')\right]\\
&& =s-\int_{-s/2}^{s/2} du \, \int_{-s/2}^{s/2} \, du'
K^2_{\infty}\left[(u-u')\right].
\end{eqnarray}
Here we used the fact that with the same precision we can put
$\rho_{\infty}(\lambda)\approx \rho_{\infty}(\lambda')\approx
\rho_{\infty}(0)$ in the above expression, and introduced the
natural scaling variables: $u=\lambda/\Delta,\, u=\lambda'/\Delta$
as well as the scaled length of the interval $s=L/\Delta$ (cf. a
similar procedure for Poissonian sequences after
Eq.(\ref{hole3})). To simplify this expression further we
introduce $u_{+}=(u+u')/2,\, r=u-u'$ as integration variables, and
use that, in fact $K_{\infty}(r)\equiv K_{\infty}(|r|)$. The
number variance takes the final form:
\begin{equation}
\Sigma_2(s)=s-\int_{-s}^{s}\,dr\,
\int_{-\frac{s}{2}+\frac{|r|}{2}}^{\frac{s}{2}-\frac{|r|}{2}}\,
du_{+} K^2_{\infty}(|r|)=s-2 \int_{0}^{s}\,dr\, (s-r)
\left[\frac{\sin{\pi r}}{\pi r}\right]^2.
\end{equation}
In fact, we are mainly interested in the large-$s$ behaviour of
this expression. To extract it, we use the identity: $2
\int_{0}^{\infty}\,dr\, \left[\frac{\sin{\pi r}}{\pi r}\right]^2=
\frac{2}{\pi} \int_{0}^{\infty}\,dx\,
\left[\frac{\sin{x}}{x}\right]^2=1$, and rewrite the above
expression as
\begin{equation}
\Sigma_2(s)=\frac{2s}{\pi} \int_{\pi s}^{\infty}\,dx\,
\frac{\sin^2{x}}{x^2}+\frac{1}{\pi^2}\int_{0}^{2\pi s}\frac{1-\cos{x}}{x}\,dx.
\end{equation}
The second integral obviously grows logarithmically with $s$
and dominates at large $s$. A more accurate evaluation gives the
asymptotic formula:
\begin{equation}
\Sigma_2(s\gg 1)=\frac{1}{\pi^2}\left[\ln{2\pi s}+\gamma+1\right]+O(1/s).
\end{equation}
where $\gamma=0.5772...$ is Euler's constant. This is much slower
than the linear growth $\Sigma_2(s\gg 1)=s$ typical for
uncorrelated (Poissonian) sequence, see Fig.~\ref{fig8}. The
explanation of the slow growth is that the sequence of eigenvalues
is, in fact, quite ordered, with quite regular spacings of the
order of $\Delta$, and therefore the number of points in the
interval does not fluctuate as much as it does for uncorrelated
sequence.

As to another important and frequently used statistical
characteristic of spectral sequences - the ``hole probability"-
its calculation amounts to investigating the asymptotics of the
Fredholm determinant of the kernel $K_{N\to\infty}$, see
Eq.(\ref{holekern}). This is a very difficult mathematical
problem, and the most elegant solution uses an advanced
mathematical technique known as the Riemann-Hilbert
method\cite{Deift}. Let us just quote the result:
\begin{equation}
A(s\gg 1)\propto \frac{1}{s^{1/4}}e^{-\frac{\pi^2}{8}s^2}.
\end{equation}
This Gaussian decay should be again contrasted with a much slower
exponential decay typical for uncorrelated sequences as indeed in
full correspondence with a ``quasiregular" structure of the random
matrix spectrum.

\subsection{ Edge scaling regime and Airy kernel}
As we already know, in the vicinity of the ``spectral edge" $x=2$
(and its counterpart $x=-2$) the Plancherel-Rotach asymptotics of
the Hermitian polynomials changes, and is basically given by the
Airy function, see Eq.(\ref{Air3}). This certainly results in
essential modifications of the large$-N$ behaviour of the mean
eigenvalue density and of the two-point kernel as long as
$|\lambda-2|\sim N^{-2/3}$. To extract the explicit formulae for
this so-called ``edge scaling" limit one may try the same strategy
as in the bulk. However, one immediately discovers that simple
substitution of Eq.(\ref{Air3}) into formula (\ref{kern777}) for
the mean density yields zero. A possible way out may be to
calculate the next-to-leading order corrections to the asymptotics
of $h_N(x)$, but we will rather follow a slightly different (and
more direct) route and consider the integral representation for
the main combination of interest:
\begin{eqnarray}\label{intrep2}
{\cal D}_N(\lambda)=h^2_{N}(\lambda)-h_{N-1}(\lambda)h_{N+1}(\lambda)=
\frac{(-1)^{N}}{2\pi}N^{2N}\,
\int_{-\infty}^{\infty}\, dq_1
\int_{-\infty}^{\infty}\,dq_2\frac{q_1-q_2}{q_1}
e^{N\left[f(q_1)+f(q_2)\right]}
\end{eqnarray}
where we exploited Eq.(\ref{intrep1},\ref{integral}), and defined,
as before, $f(q)=\ln{q}-\frac{1}{2}\left(q-i\lambda\right)^2$. To
evaluate this integral in the edge scaling limit, we follow a
familiar procedure: introduce the scaling variable
$\xi=N^{2/3}(\lambda-2)$, shift the contours of integration from
the real axis to the lines $q_{1,2}=i+\frac{t_{1,2}}{N^{1/3}}\,,
-\infty<t_{1,2}<\infty$, consider $\xi,\,t_{1,2}$ to be fixed and
finite when $N\to\infty$, and expand the integrand accordingly
around the saddle-points $t_{1,2}=0$. Simple calculation yields,
in complete analogy with Eq.(\ref{ai2}), the expression:
\begin{eqnarray}\label{ai22}
\nonumber
&& {\cal D}_N(\xi)\approx \frac{N^{2N}}{2\pi}
N^{-1/3}\,e^{N+2N^{1/3}\xi} \\
&& \times \left\{\int_{\Gamma}
\, dt_1\, e^{i\xi\,t_1+i\frac{t_1^3}{3}}\int_{\Gamma}
\, dt_2\, t_2^2\,e^{i\xi\,t_2+i\frac{t_2^3}{3}}-
\int_{\Gamma}\, dt_1\, t_1\, e^{i\xi\,t_1+i\frac{t_1^3}{3}}\int_{\Gamma}
\, dt_2\, t_2\,e^{i\xi\,t_2+i\frac{t_2^3}{3}}\right\}.
\end{eqnarray}

\begin{figure}[h!]
\centering\epsfig{file=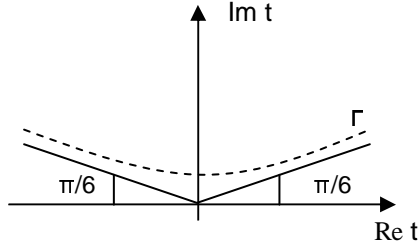} \caption{The contour of
integration $\Gamma$ in the definition of the Airy function.}
\label{fig9}\end{figure}

The only essential difference from Eq.(\ref{ai2}) which deserves
mentioning is the choice of the integration contour $\Gamma$ which
ensures the existence of all the integrals involved. Obviously,
one can not simply take $\Gamma=(-\infty,\infty)$, but a more
detailed investigation shows that the correct contour must be
chosen in such a way as to be asymptotically tangent to the line
$\mbox{Arg}{(t)}=5\pi/6$ for $\mbox{Re}\,t\to -\infty$, and
asymptotically tangent to $\mbox{Arg}{(t)}=\pi/6$ for
$\mbox{Re}\,t\to \infty$, see Fig.9. It is then evident, that
\[
Ai(\xi)=\frac{1}{\pi}\int_{\Gamma}
\, dt\, e^{i\xi\,t+i\frac{t^3}{3}},\,\,
-iAi'(\xi)=\frac{1}{\pi}\int_{\Gamma}
\, dt\,t\, e^{i\xi\,t+i\frac{t^3}{3}},\,\,
-Ai''(\xi)=\frac{1}{\pi}\int_{\Gamma}
\, dt\,t^2\, e^{i\xi\,t+i\frac{t^3}{3}}
\]
and collecting all factors we find the expression of the mean
eigenvalue density close to the ``spectral edge":
\begin{equation}\label{denedge}
\overline{\rho}(\lambda=2+\xi\,N^{-2/3})\propto \rho_{e}(\xi)=Ai'(\xi)^2-
Ai''(\xi)Ai(\xi).
\end{equation}

\begin{figure}[h!]
\centering\epsfig{file=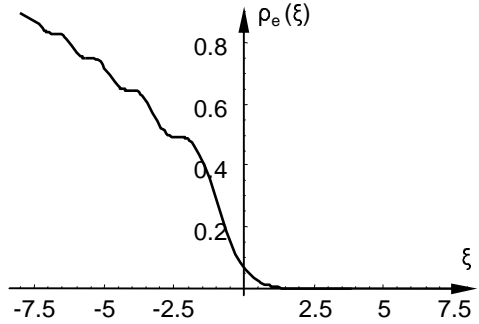} \caption{Behaviour of the
spectral density close to the spectral edge.}
\label{fig10}\end{figure}

For $\xi<0$ the function shows noticeable oscillations , see
Fig.10, with neighbouring maxima separated by distance of the
order of $\lambda_i-\lambda_{i-1}\propto \Delta_{edge}=\sim
N^{-1/3}$ and reflecting typical positions of individual
eigenvalues close to the ``spectral edge". In contrast, for
$\xi>0$ the mean density decays extremely fast, reflecting the
typical absence of the eigenvalues beyond the spectral edge.

A very similar calculation shows that under the same conditions
the kernel $K_{N}(\lambda,\lambda')$ assumes the form:
\begin{equation}\label{keredge}
K(\xi_1,\xi_2)= \frac{Ai(\xi_1)Ai'(\xi_2)-
Ai(\xi_2)Ai'(\xi_1)}{\xi_1-\xi_2}
\end{equation}
known as the {\it Airy kernel}, see \cite{Tracy}.

\section{Orthogonal polynomials versus characteristic polynomials}

Our efforts in studying  Hermite polynomials in detail were amply
rewarded by the provided possibility to arrive at the bulk and
edge scaling forms for the matrix kernel in the  corresponding
large-N limits. It is those forms which turn out to be {\it
universal}, which means independent of the particular detail of
the random matrix probability distribution, provided size of the
corresponding matrices is large enough. This is why one can hope
that the Dyson kernel would be relevant to many applications,
including properties of the Riemann $\zeta$-function. An important
issue for many years was to prove the universality for
unitary-invariant ensembles which was finally achieved, first in
\cite{PS}.

In fact, quite a few basic properties of the Hermite polynomials
are shared also by any other set of orthogonal polynomials
$\pi_{k}(x)$. Among those worth of particular mentioning is the
Christoffel-Darboux formula for the combination entering the
two-point kernel Eq.(\ref{kern1}), (cf. Eq.(\ref{darboux1})):
\begin{equation}\label{darboux11}
\sum_{k=0}^{n-1}\pi_{k}(x)\pi_{k}(y)=b_n
\frac{\pi_{n-1}(y)\pi_{n}(x)-\pi_{n-1}(x)\pi_{n}(y)}{x-y},
\end{equation}
where $b_n$ are some constants. So the problem of the universality
of the kernel (and hence, of the n-point correlation functions)
amounts to finding the appropriate large-$N$ scaling limit for the
right-hand side of Eq.(\ref{darboux11}) (in the ``bulk" of the
spectrum, or close to the spectral ``edge").

The main dissatisfaction is that explicit formulas for orthogonal
polynomials (most important, an integral representation similar to
Eq.(\ref{intrep})) are not available for general weight functions
$dw(\lambda)=e^{-Q(\lambda)}d\lambda$. For this reason we have to
devise alternative tools of constructing the orthogonal
polynomials and extracting their asymptotics. Any detailed
discussion of the relevant technique goes far beyond the modest
goals of the present set of lectures. Nevertheless, some hints
towards the essence of the powerful methods employed for that goal
will be given after a digression.

Namely, I find it instructive to discuss first a question which
seems to be quite unrelated,- the statistical properties of the
characteristic polynomials
\begin{equation} \label{Z}
Z_N(\mu)=\det\left(\mu {\bf 1}_N-\hat{H}\right)=\prod_{i=1}^N(\mu-\lambda_i)
\end{equation}
for any Hermitian matrix ensemble with invariant JPDF ${\cal
P}(\hat{H})\propto \exp\{-N\mbox{Tr}Q(\hat{H})\}$. Such objects
are very interesting on their own for many reasons. Moments of
characteristic polynomials for various types of random matrices
were much studied recently, in particular due to an attractive
possibility to use them, in a very natural way, for characterizing
``universal" features of the Riemann $\zeta$-function along the
critical line, see the pioneering paper\cite{KSn} and the lectures
by Jon Keating in this volume. The same moments also have various
interesting combinatorial interpretations, see e.g.
\cite{Strahov,DG}, and are important in applications to physics,
as I will elucidate later on.

On the other hand, addressing those moments will allow us to
arrive at the most natural way of constructing polynomials
orthogonal with respect to an arbitrary weight
$dw(\lambda)=e^{-Q(\lambda)}d\lambda$. To understand this, we
start with considering the lowest moment, which is just the
expectation value of the characteristic polynomial:
\begin{equation} \label{Z1}
E\left[Z_N(\mu)\right]=\int_{-\infty}^{\infty}dw(\lambda_1)\ldots
\int_{-\infty}^{\infty}dw(\lambda_N)
\prod_{i<j}^N(\lambda_i-\lambda_j)^2\,\prod_{i=1}^N(\mu-\lambda_i)
\end{equation}
We first notice that
\begin{equation}\label{vdm11}
\prod_{i<j}^N(\lambda_i-\lambda_j)\,\prod_{i=1}^N(\mu-\lambda_i)
\propto \det{\left(\begin{array}{cccc}1&\ldots & 1&1\\
\lambda_1&\ldots & \lambda_N& \mu\\
.&.&.&.\\ .&.&.&.\\.&.&.&.\\ \lambda^{N-1}_1&\ldots & \lambda^{N-1}_N
& \mu^{N-1}\\ \lambda^{N}_1&\ldots & \lambda^{N}_N
& \mu^{N}
\end{array}\right)}
\end{equation}
Indeed, the right-hand side is obviously a polynomial of degree
$N$ in the variable $\mu$, with roots at
$\mu=\mu_1,\mu_2,\ldots,\mu_N$. Therefore it must be of the form
$C \times \prod_{i=1}^N(\mu-\lambda_i)$, with prefactor $C$ being
a function of $\lambda_1,...,\lambda_N$. The value of such a
prefactor can be easily established by comparing both sides as
$\mu\to \infty$: the left-hand side behaves as $C\,\mu^N$, whereas
expanding the determinant with respect to the last column and
using the expression for the van der Monde determinant,
Eq.(\ref{vdm}), we see that the right-hand side grows as $\mu^N
\prod_{i<j}^N(\lambda_i-\lambda_j)$.

Exploiting Eq.(\ref{vdm11}) allows us to rewrite the expectation
value for the characteristic polynomial as
\begin{equation} \label{Z2}
E\left[Z_N(\mu)\right]\propto\int_{-\infty}^{\infty}\prod_{i=1}^N
dw(\lambda_i)
\det{\left(\begin{array}{ccc}1&\ldots & 1\\
\lambda_1&\ldots & \lambda_N\\
.&.&.\\ .&.&.\\.&.&.\\ \lambda^{N-1}_1&\ldots & \lambda^{N-1}_N
\end{array}\right)}
\det{\left(\begin{array}{cccc}1&\ldots & 1&1\\
\lambda_1&\ldots & \lambda_N& \mu\\
.&.&.&.\\ .&.&.&.\\.&.&.&.\\ \lambda^{N}_1&\ldots & \lambda^{N}_N
& \mu^{N}
\end{array}\right)},
\end{equation}
which can be further written down as the standard
sum over all permutations
$P_{\sigma}=(\sigma_1,\,\ldots,\,  \sigma_N)$
of the index set $(1,2,...,N)$:
\begin{equation} \label{Z22}
E\left[Z_N(\mu)\right]\propto\sum_{P_{\sigma}}(-1)^{|P_{\sigma}|}
\int_{-\infty}^{\infty}\prod_{i=1}^N
dw(\lambda_i)
\lambda^{0}_{\sigma_1}\ldots  \lambda^{N-1}_{\sigma_N}
\det{\left(\begin{array}{cccc}1&\ldots & 1&1\\
\lambda_1&\ldots & \lambda_N& \mu\\
.&.&.&.\\ .&.&.&.\\.&.&.&.\\ \lambda^{N}_1&\ldots & \lambda^{N}
& \mu^{N}
\end{array}\right)},
\end{equation}
where $|P_{\sigma}|=0(1)$ for even(odd) permutations. The symmetry
of the remaining determinant with respect to permutation of its
columns ensures that every term in the sum above yields exactly
the same contribution, and it is enough to consider only the first
term with $P_{\sigma}=(1,2,...,N)$, and multiply the result with
$N!$. For such a choice, the product of factors
$\lambda^{0}_{1}\ldots \lambda^{N-1}_{N}$ can be ``absorbed" in
the determinant by multiplying the $j-$th column of the latter
with the factor $\lambda^{j-1}_{j}$, for all $j=1,\ldots,N$. This
gives
\begin{equation} \label{Z4}
E\left[Z_N(\mu)\right]\propto
\int_{-\infty}^{\infty}\prod_{i=1}^N
dw(\lambda_i)
\det{\left(\begin{array}{ccccc}1&\lambda_2&\ldots & \lambda_N^{N-1}&1\\
\lambda_1&\lambda_2^2&\ldots & \lambda^N_N& \mu\\
.&.&.&.&\\ .&.&.&.&\\.&.&.&.&\\ \lambda^{N-1}_1&
\lambda^{N}_2&\ldots & \lambda^{2N-2}_N
& \mu^{N-1}\\
\lambda^{N}_1&
\lambda^{N+1}_2&\ldots & \lambda^{2N-1}_N
& \mu^{N}
\end{array}\right)}.
\end{equation}
The integral in the right-hand side is obviously a polynomial of
degree $N$ in $\mu$, which we denote $D_N(\mu)$ and write in the
final form as
\begin{equation} \label{Z5}
D_N(\mu)=
\det{\left(\begin{array}{ccccc}\int_{-\infty}^{\infty}
dw(\lambda)
&\int_{-\infty}^{\infty}
dw(\lambda)\lambda &\ldots & \int_{-\infty}^{\infty}
dw(\lambda)\lambda^{N-1}&1\\
\int_{-\infty}^{\infty}
dw(\lambda)\lambda&\int_{-\infty}^{\infty}
dw(\lambda)\lambda^2&\ldots &\int_{-\infty}^{\infty}
dw(\lambda) \lambda^N& \mu\\
.&.&.&.&\\ .&.&.&.&\\.&.&.&.&\\ \int_{-\infty}^{\infty}
dw(\lambda)\lambda^{N-1}&
\int_{-\infty}^{\infty}
dw(\lambda)\lambda^{N}&\ldots &\int_{-\infty}^{\infty}
dw(\lambda) \lambda^{2N-2}
& \mu^{N-1}\\
\int_{-\infty}^{\infty}
dw(\lambda)\lambda^{N}&
\int_{-\infty}^{\infty}
dw(\lambda)\lambda^{N+1}&\ldots & \int_{-\infty}^{\infty}
dw(\lambda)\lambda^{2N-1}
& \mu^{N}
\end{array}\right)}.
\end{equation}

The last form makes evident the following property. Multiply the
right-hand side with $dw(\mu)\mu^p$ and integrate over $\mu$. By
linearity, the factor and the integration can be ``absorbed" in
the last column of the determinant. For $p=0,1,\ldots,N-1$ this
last column will be identical to one of preceding columns, making
the whole determinant vanishing, so that
\begin{equation}\label{z6}
\int_{-\infty}^{\infty} dw(\mu)\mu^p D_N(\mu)=0,\quad
p=0,1,\ldots,N-1.
\end{equation}

Moreover, it is easy to satisfy oneself that the polynomial
$D_N(\mu)$ can be written as $D_N(\mu)=D_{N-1}\mu^N+\ldots $,
where the leading coefficient
$D_{N-1}=\det{\left(\int_{-\infty}^{\infty}
dw(\lambda)\lambda^{i+j}\right)_{i,j=0}^N}$ is necessarily
positive: $D_{n-1}>0$. The last fact immediately follows from the
positivity of the quadratic form:
\[
G(x_1,\ldots,x_N)=\int_{-\infty}^{\infty}
dw(\lambda)\left(\sum_{i=1}^N\,x_i\lambda^{i}\right)^2=
\sum_{i,j}^Nx_ix_j\int_{-\infty}^{\infty}
dw(\lambda)\lambda^{i+j}
\]
Finally, notice that
\begin{equation}
\int_{-\infty}^{\infty}
dw(\mu)D_N^2(\mu)=\int_{-\infty}^{\infty}
dw(\mu)D_N(\mu)\left[D_{N-1}\mu^N+lower\,\, powers\right]
\end{equation}
\[
=D_{N-1}\int_{-\infty}^{\infty}
dw(\mu)D_N(\mu)\mu^N=D_{N-1}D_N
\]
where we first exploited Eq.(\ref{z6}) and at the last stage
Eq.(\ref{Z5}). Combining all these facts together we thus proved
that the polynomials
$\pi_N(\lambda)=\frac{1}{\sqrt{D_{N-1}D_N}}D(\lambda)$ form the
orthogonal (and normalized to unity) set with respect to the given
measure $dw(\lambda)$. Moreover, our discussion makes it
immediately clear that the expectation value of the {\it
characteristic} polynomial $Z_N(\mu)$ for any given random matrix
ensemble is nothing else, but just the corresponding {\it monic}
orthogonal polynomial:
\begin{equation}\label{z8}
E\left[Z_N(\mu)\right]=\pi^{(m)}_{N}(\mu),
\end{equation}
whose leading coefficient is unity. Leaving aside the modern
random matrix interpretation the combination of the right hand
sides of the formulas Eq.(\ref{z8}) and Eq.(\ref{Z1}) goes back,
according to \cite{Szego}, to Heine-Borel work of 1878, and as
such is completely classical.

The random matrix interpretation is however quite instructive,
since it suggests to consider also higher moments of the
characteristic polynomials, and even more general objects like the
correlation functions
\begin{equation}\label{corrf}
{\cal C}_k(\mu_1,\mu_2,\ldots,\mu_k)=E\left[Z_N(\mu_1)Z_N(\mu_2)
\ldots Z_N(\mu_k)\right].\end{equation} Let us start with
considering
\begin{equation}\label{corrf1}
{\cal C}_2(\mu_1,\mu_2)=
\int_{-\infty}^{\infty}dw(\lambda_1)\ldots
\int_{-\infty}^{\infty}dw(\lambda_N)
\prod_{i<j}^N(\lambda_i-\lambda_j)^2\,\prod_{i=1}^N(\mu_1-\lambda_i)
\prod_{i=1}^N(\mu_2-\lambda_i).
\end{equation}
Using the notation $\Delta_N(\lambda_1,\ldots,\lambda_N)$ for the van
der Monde determinant, see Eq.(\ref{vdm}), we further notice that
\[
\Delta_{N+2}(\lambda_1,\ldots,\lambda_N,\mu_1,\mu_2)=
\Delta_N(\lambda_1,\ldots,\lambda_N)\times (\mu_1-\mu_2)
\prod_{i=1}^N(\mu_1-\lambda_i) \prod_{i=1}^N(\mu_2-\lambda_i),
\]
which allows us to rewrite the correlation function as
\[
{\cal C}_2(\mu_1,\mu_2)=
\frac{1}{(\mu_1-\mu_2)}\int_{-\infty}^{\infty}\prod^N_{i=1}
dw(\lambda_i)\,\Delta_{N}(\lambda_1,\ldots,\lambda_N)
\Delta_{N+2}(\lambda_1,\ldots,\lambda_N,\mu_1,\mu_2).
\]
Now we replace each entry $\lambda_i^j$ in both van der Monde
determinant  factors with the orthogonal polynomial
$\pi_j(\lambda_i)$ (cf. eq.(\ref{vdm1})), and further expand the
first factor as a sum over permutations:
$\Delta_{N}(\lambda_1,\ldots,\lambda_N)\propto \sum_{P}(-1)^{|P|}
\pi_{0}(\lambda_{\sigma_1})\ldots  \pi_{N-1}(\lambda_{\sigma_N})$.
Further using permutational symmetry of the second determinant, we
again see that every term yields after integration the same
contribution. Up to a proportionality factor we can therefore
rewrite the correlation function as
\begin{eqnarray}
{\cal C}_2(\mu_1,\mu_2)&=&
\frac{1}{(\mu_1-\mu_2)}\int_{-\infty}^{\infty}\prod^N_{i=1}
dw(\lambda_i)\,\pi_{0}(\lambda_{1})\ldots  \pi_{N-1}(\lambda_{N})\\
\nonumber
&\times&\det{\left(\begin{array}{ccccc}\pi_{0}(\lambda_{1})&
\pi_{0}(\lambda_2)&\ldots &\pi_{0}(\mu_1)&\pi_{0}(\mu_2)\\
\pi_{1}(\lambda_1)&\pi_{1}(\lambda_2)&\ldots &
\pi_{1}(\mu_1)& \pi_{1}(\mu_2) \\
.&.&.&.&\\ .&.&.&.&\\.&.&.&.&\\ \pi_{N}(\lambda_1)&
\pi_{N}(\lambda_2)&\ldots & \pi_{N}(\mu_1)
& \pi_{N}(\mu_2)\\
\pi_{N+1}(\lambda_1)&
\pi_{N+1}(\lambda_2)&\ldots &\pi_{N+1}(\mu_1)
& \pi_{N+1}(\mu_2)
\end{array}\right)}.
\end{eqnarray}
At the next step we absorb the factors $
\pi_{0}(\lambda_{1}),\dots, \pi_{N-1}(\lambda_N)$ inside the
determinant by multiplying the first column with
$\pi_{0}(\lambda_{1})$,..., the $N-th$ column with
$\pi_{N-1}(\lambda_N)$, and leaving the last two columns intact.
By linearity, we can also absorb the product of the integrals
inside the determinant by integrating the first column over
$\lambda_1$,..., and $N-th$ column over $\lambda_N$. Due to the
orthogonality, the first $N$ columns of the resulting determinant
after integration contain zero components off-diagonal, whereas
the entries on the main diagonal are equal to the normalization
constants $c_k=\int_{-\infty}^{\infty}
dw(\lambda)\,\pi^2_{k}(\lambda),\,\, k=0,\dots,N$. Therefore, the
resulting determinant is easy to calculate and, up to a
multiplicative constant we arrive to the following simple formula:
\begin{equation}\label{corrf2}
{\cal C}_2(\mu_1,\mu_2)\propto \frac{1}{(\mu_1-\mu_2)}
\det{\left(\begin{array}{cc} \pi_{N}(\mu_1) & \pi_{N}(\mu_2)\\
\pi_{N+1}(\mu_1) & \pi_{N+1}(\mu_2)\end{array}\right)}.
\end{equation}
In particular, for the second moment of the characteristic
polynomial we have the expression
\begin{equation}\label{mom2}
E[Z^2(\mu)]=\lim_{\mu_1\to\mu_2=\mu} {\cal
C}_2(\mu_1,\mu_2)\propto \det{\left(\begin{array}{cc} \pi_{N}(\mu)
& \pi'_{N}(\mu)\\ \pi_{N+1}(\mu) &
\pi'_{N+1}(\mu)\end{array}\right)}.
\end{equation}

This procedure can be very straightforwardly extended to higher
order correlation functions\cite{BH,MN}, and higher order
moments\cite{FW} of the characteristic polynomials. The general
structure is always the same, and is given in the form of a
determinant whose entries are orthogonal polynomials of increasing
order.

One more observation deserving mentioning here is that the
structure of the two-point correlation function of characteristic
polynomials is identical to that of the Christoffel-Darboux, which
is the main building block of the kernel function,
Eq.(\ref{kern1}). Moreover, comparing the above formula
(\ref{mom2}) for the gaussian case with expressions
(\ref{kern77},\ref{darboux3}), one notices a great degree in
similarity between the structure of mean eigenvalue density and
that for the second moment of the characteristic polynomial. All
these similarities are not accidental, and there exists a general
relation between the two types of quantities as I proceed to
demonstrate on the simplest example. For this we recall that the
mean eigenvalue density $\overline{\rho_N(\lambda)}$ is just the
one-point correlation function, see Eq.(\ref{24}), and according
to Eq.(\ref{21}) and Eq.(\ref{JPDG}) can be written as
\begin{eqnarray}
\nonumber {\cal R}_{1}(\lambda) &=&N \int \,{\cal
P}(\lambda,\lambda_2,\ldots,\lambda_N) \,\,
d\lambda_{2}\,\ldots\,\lambda_N\\ &\propto& e^{-Q(\lambda)} \int
d\lambda_{2}\,\ldots\,\lambda_N e^{-\sum_{i=2}^NQ(\lambda_i)}
\prod_{i=2}^N(\lambda-\lambda_i)^2 \prod_{2\le i<j\le
N}\left(\lambda_i-\lambda_j\right)^2. \label{211}
\end{eqnarray}
It is immediately evident after simple renumbering
$(\lambda_2,\ldots,\lambda_N)\to (\lambda_1,\ldots,\lambda_{N-1})$
that the integral in the second line allows a clear interpretation
as the second moment of the characteristic polynomial
$E[Z^2_{N-1}(\lambda)]$ of a random matrix $H_{N-1}$ distributed
according to the same joint probability density function ${\cal
P}\left(H_{N-1}\right) d\hat{H}_{N-1},$ but of reduced size $N-1$,
see Eq.(\ref{Z1}) for comparison. We therefore have a general
relation between the mean eigenvalue density and the second moment
of the characteristic polynomial of the reduced-size matrix:
\begin{equation}
\overline{\rho_N(\lambda)}\propto e^{-Q(\lambda)}
\overline{\left[\det{\left(\lambda {\bf 1}_N-\hat{H}_{N-1}\right)}
\right]^2}
\end{equation}
which explains the observed similarity. This type of relations,
and their natural generalizations to higher-order correlation
functions hold for general invariant ensembles and were found
helpful in several applications; e.g. for the so-called ``chiral"
ensembles (notion of such ensembles is shortly discussed in the
very end of these notes) in \cite{AK}, for non-Hermitian matrices
with complex eigenvalues see examples and further references in
\cite{FS}); for real symmetric matrices see the recent
paper\cite{fluc}.

Now let us discuss another important class of correlation
functions involving characteristic polynomials, - namely one
combining both positive and negative moments, the simplest example
being the expectation value of the ratio:
\begin{equation}\label{ratio}
{\cal K}_{N}(\mu,\nu)= E\left[ \frac{ Z_N(\mu)}{Z_N(\nu)}\right].
\end{equation}
For such an object to be well-defined it is necessary to
regularize the characteristic polynomial in the denominator
$Z_N(\nu)=\det\left(\nu {\bf 1}_N-\hat{H}\right)$ by considering
the complex-valued spectral parameter $\nu$ such that
$\mbox{Im}\nu\ne 0$. Further generalizations include more than one
polynomial in numerator and/or denominator.

Such  objects turned out to be indispensable tools in applications
of random matrices to physical problems. In fact, in all
applications a very fundamental role is played by the resolvent
matrix $(\mu{\bf 1}_N- \hat{H})^{-1}$, and statistics of its
entries is of great interest. In particular, the familiar
eigenvalue density $\rho(\nu)$ can be extracted from the trace of
the resolvent as
\begin{equation}\label{rho}
\rho(\nu)=\frac{1}{\pi}\lim_{\mbox{Im}\mu\to 0^{-}}
\mbox{Im}\mbox{Tr}\frac{1}{\mu{\bf 1}_N- \hat{H}}.
\end{equation}

It is easy to understand that one can get access to such an
object, and more general correlation functions of the traces of
the resolvent by using the identity:
\begin{equation}\label{rho1}
\mbox{Tr}\frac{1}{\mu{\bf 1}_N- \hat{H}}=-\frac{\partial}{\partial
\nu} \frac{Z_N(\mu)}{Z_N(\nu)}|_{\mu=\nu}.
\end{equation}
We conclude that the products of ratios of characteristic
polynomials can be used to extract the multipoint correlation
function of spectral densities (see an example below). Moreover,
distributions of some other interesting quantities as, e.g.
individual entries of the resolvent, or statistics of eigenvalues
as functions of some parameter can be characterized in terms of
general correlation functions of ratios, see \cite{AS} for more
details and examples. Thus, that type of the correlation function
is even more informative than one containing products of only
positive moments of the characteristic polynomials.

In fact, it turns out that there exists a general relation between
the two types of the correlation functions, which is discussed in
full generality in recent papers \cite{FS1,SF2,BDS,BS}. Here we
would like to illustrate such a relation on the simplest example,
Eq.(\ref{ratio}). To this end let us use the following identity:
\begin{equation}\label{ident}
\left[Z_N(\nu)\right]^{-1}=\frac{1}{\prod_{i=1}^N{(\nu-\lambda_i)}} =
\sum_{k=1}^N\frac{1}{\nu-\lambda_k}\prod_{i\ne k}^N
\frac{1}{\lambda_i-\lambda_k}
\end{equation}
and integrate the ratio of the two characteristic polynomials over
the joint probability density of all the eigenvalues. When
performing integrations, each of $N$ terms in the sum in
Eq.(\ref{ident}) produces identical contributions, so that we can
take one term with $k=1$ and multiply the result by $N$.
Representing $\Delta^2(\lambda_1,\ldots, \lambda_N)=\prod_{2\le
i}(\lambda_1-\lambda_i)^2 \prod_{2\le i<j}
(\lambda_i-\lambda_j)^2$, and observing some cancellations, we
have
\begin{eqnarray}
\nonumber && {\cal K}_{N}(\mu,\nu)\propto \int
dw(\lambda_1) \frac{\mu-\lambda_1}{\nu-\lambda_1} \int
dw(\lambda_2)\, ...\, dw(\lambda_N)
\prod_{2\le i<j}^N(\lambda_i-\lambda_j)^2
\prod_{i=2}^N{(\lambda_1-\lambda_i)}{(\mu-\lambda_i)}
\\ &&\propto \int
dw(\lambda_1)  \frac{\mu-\lambda_1}{\nu-\lambda_1} \times
\overline{\mbox{det}\left(\lambda_1-\hat{H}_{N-1}\right)\mbox{det}
\left(\mu-\hat{H}_{N-1}\right)}. \label{Herspe}
\end{eqnarray}
The average value of the products of two characteristic
polynomials found by us in Eq.(\ref{corrf2}) can now be inserted
into the integral entering Eq.(\ref{Herspe}), and the resulting
expression can be again written in the form of a $2\times 2$
determinant:
\begin{equation}\label{Her44}
 K_{N}(\mu,\nu)\propto
\mbox{det}\left(\begin{array}{cc}
 \pi_{N-1}\left(\mu\right)&
f_{N-1}(\nu)
\\  \pi_{N}\left(\mu \right)&
f_N(\nu)
\end{array}\right)
\end{equation}
where $f_N(\nu)$ stands for the so-called Cauchy transform of the
orthogonal polynomial
\begin{equation}\label{fdef}
f_N(\nu)=\frac{1}{2\pi i}
\int^{\infty}_{-\infty}\frac{dw(\lambda)}{{\nu-\lambda}}
\,\pi_{N}\left(\lambda\right)
\end{equation}

The emerged functions $f_{N}(\nu)$ is a rather new feature in
Random Matrix Theory. It is instructive to have a closer look at
their properties for the simplest case of the Gaussian Ensemble,
$Q(\lambda)=N\lambda^2/2$. It turns out that for such a case the
functions $f_N(\nu)$ are, in fact, related to the so-called
generalized Hermite functions ${\cal H}_{N}$ which are second
-non-polynomial- solutions of the same differential equation which
is satisfied by Hermite polynomials themselves. The functions also
have a convenient integral representations, which can be obtained
in the most straightforward way by substituting the identity
\[
\frac{1}{\nu-\lambda}\propto\int_0^{\infty} dt
e^{it\mbox{\small sgn}{\left[\mbox{\small Im}\nu\right]}(\nu-\lambda)}
\]
into the definition (\ref{fdef}), replacing the Hermite polynomial
with its integral representation, Eq.(\ref{intrep}), exchanging
the order of integrations and performing the $\lambda-$integral
explicitly. Such a procedure results in
\[
f_{N+n}(\nu)\propto \int_0^{\infty} dt t^{N+n} e^{-N
\left(\frac{t^2}{2}-it\mbox{\small sgn}{\left[ \mbox{\small
Im}\nu\right]}\nu\right)}.
\]
Note, that this is precisely the integral (\ref{integral}) whose
large-N asymptotics for real $\nu$ we studied in the course of our
saddle-point analysis. The results can be immediately extended to
complex $\nu$, and in the ``bulk scaling" limit we arrive to the
following asymptotics of the correlation function (\ref{Her44})
close to the origin
\begin{equation}\label{FS}
\lim_{N\to\infty} {\cal K}_N(\mu,\nu)={\cal K}_{\infty}\left[
N\rho_{\infty}(0)(\mu-\nu)\right],\quad
K_{\infty}(r)\propto\left\{\begin{array}{cc}
e^{-i\pi r}& \mbox{if Im}(\nu)>0 \\
e^{i\pi r} & \mbox{if Im}(\nu)>0 \end{array}\right..
\end{equation}
In a similar, although more elaborate way one can calculate an
arbitrary correlation function containing ratios and products of
characteristic polynomials \cite{FS1,BDS,BS}. The detailed
analysis shows that the kernel $ S(\mu,\nu)={\cal
K}(\mu,\nu)/(\mu-\nu)$ and its scaling form
$S_{\infty}(r)\propto\frac{K_{\infty}(r)}{r}$ play the role of a
building block for more general correlation functions involving
ratios, in the same way as the Dyson kernel (\ref{Dyson}) plays
similar role for the n-point correlation functions of eigenvalue
densities. This is a new type of ``kernel function" with structure
different from the standard random matrix kernel Eq.(\ref{kern}).
The third type of such kernels - made from functions $f_N(\nu)$
alone - arises when considering only negative moments of the
characteristic polynomials.

To give an instructive example of the form emerging consider
\begin{equation}\label{ratio2}
{\cal K}_N(\mu_1,\mu_2,\nu_1,\nu_2)=E\left[
\frac{ Z_N(\mu_1)}{Z_N(\nu_1)}\frac{ Z_N(\mu_2)}{Z_N(\nu_2)}\right]=
\frac{(\mu_1-\nu_1)(\mu_1-\nu_2)
(\mu_2-\nu_1)(\mu_2-\nu_2)}{(\mu_1-\mu_2)(\nu_1-\nu_2)}
\end{equation}
\[\times\mbox{det}\left(\begin{array}{cc}
 S\left(\mu_1,\nu_1\right)&
S\left(\mu_1,\nu_2\right)
\\  S\left(\mu_2,\nu_1\right)&
S\left(\mu_2,\nu_2\right)
\end{array}\right).
\]

Assuming Im$\,\nu_1>0$, Im$\,\nu_2<0$, both infinitesimal, we find
in the bulk scaling limit such that both
$N\rho(0)\mu_{1,2}=\zeta_{1,2}$ and
$N\rho(0)\nu_{1,2}=\kappa_{1,2}$ are finite the following
expression (see e.g. \cite{SF2}, or \cite{AS})
\begin{equation}\label{ratio3}
\lim_{N\to\infty}{\cal
K}_N(\mu_1,\mu_2,\nu_1,\nu_2)=
{\cal K}_{\infty}(\zeta_1,\zeta_2,\kappa_1,\kappa_2)
\end{equation}
\begin{equation}
=\frac{e^{i\pi(\zeta_1-\zeta_2)}}{\zeta_1-\zeta_2}
\left[e^{i\pi(\kappa_1-\kappa_2)}\frac{(\kappa_1-\zeta_1)
(\kappa_2-\zeta_2)}
{\kappa_1-\kappa_2}-e^{-i\pi(\kappa_1-\kappa_2)}
\frac{(\kappa_1-\zeta_2)(\kappa_2-\zeta_1
)}{\kappa_1-\kappa_2}\right]. \nonumber
\end{equation}

This formula can be further utilized for many goals. For example,
it is a useful exercise to understand how the scaling limit of the
two-point cluster function (\ref{kern8}) can be extracted from
such an expression (hint: the cluster function is related to the
correlation function of eigenvalue densities by Eq.(\ref{2p});
exploit the relations (\ref{rho}),(\ref{rho1})).

All these developments, - important and interesting on their own,
indirectly prepared the ground for discussing the mathematical
framework for a proof of universality in the large-$N$ limit. As
was already mentioned, the main obstacle was the absence of any
sensible integral representation for general orthogonal
polynomials and their Cauchy transforms. The method which
circumvents this obstacle in the most elegant fashion is based on
the possibility to define {\it both} orthogonal polynomials {\it
and} their Cauchy transforms in a way proposed by Fokas, Its and
Kitaev, see references in \cite{Deift}, as elements of a (matrix
valued) solution of the following (Riemann-Hilbert) problem. The
latter can be introduced as follows. Let the contour $\Sigma$ be
the real axis orientated from the left to the right. The upper
half of the complex plane with respect to the contour will be
called the positive one and the lower half - the negative one. Fix
an integer $n\geq 0$ and the measure $w(z)=e^{-Q(z)}$ and define
the Riemann-Hilbert problem as that of finding a $2\times 2$
matrix valued function $Y=Y^{(n)}(z)$ satisfying the following
conditions:
\begin{itemize}
  \item $Y^{(n)}(z)- \mbox{analytic}\;\mbox{in}\;
   \textsc{C}\setminus\Sigma $
  \item $Y^{(n)}_{+}(z)=Y^{(n)}_{-}(z)\left(
  \begin{array}{cc}
    1 & w(z) \\
    0 & 1 \
  \end{array}
  \right),\;z\in \Sigma$
  \item $Y^{(n)}(z)\mapsto\left(I+{\mathcal{O}}(z^{-1})\right)
  \left(\begin{array}{cc}
    z^n & 0 \\
    0 & z^{-n} \
  \end{array}
  \right)\;\; \mbox{as}\;\; z\mapsto \infty $
\end{itemize}
Here  $Y^{(n)}_{\pm}(z)$ denotes the limit of  $Y^{(n)}(z')$ as
$z'\mapsto z\in \Sigma$ from the positive/negative side of the
complex plane. It may be proved (see \cite{Deift}) that the
solution of such a problem is unique and is given by
\begin{equation}\label{R-H Solution}
Y^{(n)}(z)=\left(
\begin{array}{cc}
  \pi_n(z) & f_n(z) \\
  \gamma_{n-1}\pi_{n-1}(z) & \gamma_{n-1}f_{n-1}(z)
\end{array}
\right),\;\;\;\mbox{Im}\;z\neq 0
\end{equation}
where the constants $\gamma_n$ are simply related to the
normalization of the corresponding polynomials: $\gamma_n=-2\pi i
[\int_{-\infty}^{\infty} dw \pi^2_n]^{-1}$.

On comparing formulae (\ref{Her44}) and (\ref{R-H Solution}) we
observe that the structure of the correlation function
${\mathcal{K}}_{N}(\mu,\nu)$ is very intimately related to the
above Riemann-Hilbert problem. In fact, for $\mu=\nu=z$ the
matrices involved are identical (even the constant $\gamma_{n-1}$
in Eq.(\ref{Her44}) emerges when we replace $\propto$ with exact
equality sign). Actually, all three types of kernels can be
expressed in terms of the solution of the Riemann-Hilbert problem.
The original works \cite{Deift,BI} dealt only with the standard
kernel built from polynomials alone. From that point of view the
presence of Cauchy transforms in the Riemann-Hilbert problem might
seem to be quite mysterious, and even superfluous. Now, after
revealing the role and the meaning of more general kernels the
picture can be considered complete, and the presence of the Cauchy
transforms has its logical justification.

The relation to the Riemann-Hilbert problem is the starting point
for a very efficient method of extracting the large$-N$
asymptotics  for essentially any potential function $Q(x)$
entering the probability distribution measure.  The corresponding
machinery is known as the variant of the steepest
descent/stationary phase method introduced for Riemann-Hilbert
problems by Deift and Zhou. It is discussed at length in the book
by Deift\cite{Deift} which can be recommended to the interested
reader for further details. In this way the universality was
verified for all three types of kernels pertinent to the random
matrix theory not only for bulk of the spectrum\cite{SF2}, but
also for the spectral edges . In our considerations of the
Gaussian Unitary Ensemble we already encountered the edge scaling
regime where the spectral properties were parameterized by the
Airy functions $Ai(x)$. Dealing with ratios of characteristic
polynomials in such a regime requires second solution of the Airy
equation denoted by $Bi(x)$, see \cite{AF}.

We finish our exposition by claiming that there exist other
interesting classes of matrix ensembles which attracted a
considerable attention recently, see the paper\cite{sycl} for more
detail on the classification of random matrices by underlying
symmetries. In the present framework we only mention one of them -
the so-called {\it chiral} GUE. The corresponding $2N\times 2N$
matrices are of the form
$\hat{H}_{ch}=\left(\begin{array}{cc}{\bf 0}_N&\hat{J}\\
\hat{J}^{\dagger}& {\bf 0}_N \end{array}\right)$, where $\hat{J}$
of a general complex matrix. They were introduced to provide a
background for calculating  the universal part of the microscopic
level density for the Euclidian QCD Dirac operator, see \cite{Ver}
and references therein, and also have relevance for applications
to condensed matter physics. The eigenvalues of such matrices
appear in pairs $\pm\lambda _k\,\,,\,\, k=1,...,N$. It is easy to
understand that the origin $\lambda=0$ plays a specific role in
such matrices, and close to this point eigenvalue correlations are
rather different from those of the GUE, and described by the
so-called Bessel kernels\cite{T2}. An alternative way of looking
essentially at the same problem is to consider the random matrices
of Wishart type $\hat{W}= \hat{J}^{\dagger}\hat{J}$, where the
role of the special point is again played by the origin (in such
context the origin is frequently referred to as the ``hard
spectral edge", since no eigenvalues are possible beyond that
point. This should be contrasted with the Airy regime close to the
semicircle edge, the latter being sometimes referred to as the
"soft edge" of the spectrum. ). The corresponding problems for
products and ratios of characteristic polynomials were treated in
full rigor by Riemann-Hilbert technique by
Vanlessen\cite{vanlessen}, and in a less formal way in \cite{AF}.

\subsection{Acknowledgement}

My own understanding in the topics covered was shaped in the
course of numerous discussions and valuable contacts with many
colleagues. I am particularly grateful to Gernot Akemann, Jon
Keating, Boris Khoruzhenko and Eugene Strahov for fruitful
collaborations on various facets of the Random Matrix theory which
I enjoyed in recent years. I am indebted to Nina Snaith and
Francesco Mezzadri for their invitation to give this lecture
course, and for careful editing of the notes. It allowed me to
spend a few wonderful weeks in the most pleasant and stimulating
atmosphere of the Newton Institute, Cambridge whose hospitality
and financial support I acknowledge with thanks. Finally, it is my
pleasure to thank Guler Ergun for her assistance with preparing
this manuscript for publication.

\end{document}